\documentclass[aip,numerical,amssymb,amsmath,pof]{revtex4-1}
\usepackage{graphicx}
\usepackage{xcolor}
\usepackage{bm}%

\usepackage{amsmath,amsfonts,amssymb}
\usepackage{times}

\def\rot{\mathop{\rm rot}\nolimits}
\def\div{\mathop{\rm div}\nolimits} 
\def\gsim{\lower.4ex\hbox{$\;\buildrel >\over{\scriptstyle\sim}\;$}} 
\def\lsim{\lower.4ex\hbox{$\;\buildrel <\over{\scriptstyle\sim}\;$}} 
\def\rin{r_{\rm in}}

\newcommand{\Rey}{\mathrm{Re}}

\newcommand{\Rm}{\mathrm{Rm}}

\newcommand{\Rin}{R_\mathrm{in}}
\newcommand{\Rout}{R_\mathrm{out}}
\newcommand{\Ro}{\mathrm{Ro}}

\newcommand{\muh}{{\mu}}

\newcommand{\Pm}{\mathrm{Pm}}
\newcommand{\Ha}{\mathrm{Ha}}
\newcommand{\Mm}{\mathrm{Mm}}

\newcommand{\mperm}{\mu_0}
\newcommand{\mdiff}{\eta}

\def\rot{\mathop{\rm curl}\nolimits}
\def\div{\mathop{\rm div}\nolimits} 
\def\gsim{\lower.4ex\hbox{$\;\buildrel >\over{\scriptstyle\sim}\;$}} 
\def\lsim{\lower.4ex\hbox{$\;\buildrel <\over{\scriptstyle\sim}\;$}} 
\def\q{\qquad}
\def\beg{\begin{eqnarray}}
\def\ende{\end{eqnarray}}
\renewcommand{\vec}[1]{\mbox{\boldmath $#1$}}
\newcommand{\etah}{r_{\rm in}}
\newcommand{\Om}{{\it \Omega}}
\def\A{Alfv\'en}
\def\R{R\"udiger}

\begin{document}
\title{Subcritical excitation of the current-driven Tayler instability by super-rotation}
\author{G. \R} \email{gruediger@aip.de} 
\affiliation{Leibniz-Institut f\"ur Astrophysik Potsdam, An der Sternwarte, 14482 Potsdam, Germany}
\affiliation{Helmholtz-Zentrum Dresden-Rossendorf, P.O. Box 510119, D-01314 Dresden, Germany}

\author{M. Schultz}
\affiliation{Leibniz-Institut f\"ur Astrophysik Potsdam, An der Sternwarte, 14482 Potsdam, Germany}
\affiliation{Helmholtz-Zentrum Dresden-Rossendorf, P.O. Box 510119, D-01314 Dresden, Germany}
\author{M. Gellert}
\affiliation{Leibniz-Institut f\"ur Astrophysik Potsdam, An der Sternwarte, 14482 Potsdam, Germany}
\author{F. Stefani} 
\affiliation{Helmholtz-Zentrum Dresden-Rossendorf, P.O. Box 510119, D-01314 Dresden, Germany}

\date{\today}

\begin{abstract}
It is known that in a hydrodynamic Taylor-Couette system uniform rotation or a rotation law  with positive shear (`super-rotation') are linearly stable. It is also known  that  a  conducting fluid under the presence  of a sufficiently strong axial electric-current   becomes unstable against nonaxisymmetric disturbances. It is thus suggestive  that   a cylindric  pinch formed  by a homogeneous axial electric-current is stabilized by rotation laws with  ${\rm d}\Om/{\rm d} R\geq 0$.  However, for magnetic Prandtl number  $\Pm\neq 1$ and for slow rotation also rigid rotation and super-rotation   {\em support} the instability by lowering their critical Hartmann numbers.   For  super-rotation in narrow gaps and for modest rotation rates this double-diffusive instability even  exists  for toroidal magnetic fields with rather arbitrary radial profiles, the current-free profile $B_\phi\propto 1/R$ included.  -- 
 For rigid rotation and for super-rotation  the sign of the azimuthal drift of the nonaxisymmetric hydromagnetic instability pattern strongly  depends on the magnetic Prandtl number. The pattern counterrotates with the flow for $\Pm \ll 1$ and it corotates  for $\Pm \gg 1$ while  for rotation laws with negative shear  the instability  pattern migrates in the direction of the basic rotation  for all $\Pm$. 
  
  An axial electric-current of minimal 3.6 kAmp flowing inside or outside the inner cylinder  suffices to realize the double-diffusive  instability  for super-rotation  in experiments using liquid sodium as the conducting fluid between the rotating cylinders. The limit is 11 kAmp if a gallium alloy is used.
\end{abstract}

\maketitle

\sloppy

 \section{Introduction}
 A well-known  instability of toroidal fields is the magneto hydrodynamical pinch-type current-driven  instability  which is basically nonaxisymmetric\cite{T73}. The toroidal field becomes unstable if  a certain magnetic field amplitude is exceeded  depending on the radial profile of the field which forms the electric-current pattern.  It  is also known that for unity magnetic Prandtl number a global rotation of the system increases  the critical field amplitude. The latter is strongly reduced, however,  if the rotation decreases outwards (i.e. ${\rm d}\Om/{\rm d} R<0$,  `sub-rotation'). The formal reason is  that sub-rotation   becomes (Rayleigh-) unstable even  in the hydrodynamic regime if  it is steep enough. More important is the existence of a nonaxisymmetric instability for such rotation laws even for current-free toroidal fields ($B_\phi\propto 1/R$) which we have called  azimuthal magnetorotational instability (AMRI\cite{R07,HTR09,R14}). It appears   for all values  of the magnetic Prandtl number for rather low Hartmann numbers but for  large magnetic Reynolds numbers of the basic rotation.  This phenomenon  also explains the general destabilization of toroidal fields  by rotation laws with $\Om$  {\em decreasing} outwards.

The question arises about the role of `super-rotation', i.e. rotation laws with ${\rm d}\Om/{\rm d} R>0$, which are  linearly  stable in the hydrodynamic regime \cite{T36,SG59}.   
The nonlinear behavior is less clear as  some Taylor-Couette experiments have  shown instability 
 in this regime\cite{BST10,BC12}. Superrotation 
cannot be destabilized by  the standard  magnetorotational 
instability with axial external fields.  Inspired by the discovery of the axisymmetric helical MRI 
 a WKB method for inviscid  fluids in current-free helical background fields has been  applied  providing
two  limits of instability  in terms 
of the shear in   the rotation law\cite{L06}. 
An upper threshold 
suggests  a magnetic destabilization of
super-rotating flows  for very strong positive shear. 
 A similar phenomenon  has been reported by Bonanno \& Urpin (2008) resulting from a local analysis for a helical field under the influence of super-rotation\cite{BU08}.
 Later 
it has been shown with a dispersion relation  for inductionless fluids (see the Appendix) 
that the   stability curve does not cross the line representing the differentially rotating pinch formed by  uniform electric-current  suggesting  instability for both signs of shear\cite{KS13}.

By means of a corresponding approximation Acheson (1978) showed  that for  fast-rotation the  current-driven instability of toroidal fields may be stabilized by positive shear \cite{A78}. If this is true we expect in the solar low latitudes where in the bulk of the convection zone the equatorial $\Om$ increases outwards that  the toroidal field is stabilized and can be amplified to much higher values than it would be true for the opposite rotation law.  Contrary to that a rotation law with negative shear -- as it exists in higher solar latitudes -- strongly destabilizes the fields so that they cannot reach high amplitudes.  It is shown here by use of a simplifying  cylinder geometry that indeed for not too small magnetic Prandtl numbers super-rotation stabilizes toroidal magnetic fields while  sub-rotation strongly destabilizes toroidal magnetic fields. On the other hand,
small magnetic Prandtl number and  slow rotation  of any  rotation law -- including rigid rotation --  lead to  {\em lower} critical magnetic field strengths than needed for destabilization at  $\Om=0$. We shall show in the present paper that for  ${\rm d}\Om/{\rm d} R\geq 0$ and for slow rotation the relaxation of the excitation conditions compared with the resting container belongs to the double-diffusive phenomena which disappear { if the the molecular viscosity equals the molecular resistivity}.

A Taylor-Couette container is considered which confines  a  toroidal magnetic field with  amplitudes fixed  at the cylinders which may rotate with different rotation rates. The gap between the cylinders is considered as variable. Normalized with the outer radius  $\Rout$ the inner radius $\Rin$ is $\geq 0.5$. The cylinders are unbounded in axial direction.

The fluid  between the cylinders is assumed to be incompressible and
 dissipative { with the kinematic viscosity $\nu$ and the magnetic diffusivity $\eta$}. Derived from  the conservation of angular momentum the rotation law
 $\Om(R)$ in the fluid is
 \beg
    \Om(R)=a+\frac{b}{R^2}
 \label{Om}
 \ende
 with
 \beg
  a=\frac{\muh-\etah^2}{1-\etah^2} \Om_{\rm in}, \q 
  b=\frac{1-\muh}{1-\etah^2}\Rin^2 \Om_{\rm in},
 \ende
 where
\beg
\rin=\frac{R_{\rm{in}}}{R_{\rm{out}}}, \q\q\q
\mu=\frac{\Om_{\rm{out}}}{\Om_{\rm{in}}}.
\label{mu}
\ende
$\Om_{\rm in}$ and $\Om_{\rm out}$ are the imposed rotation rates of
the inner and outer cylinders.
  After the Rayleigh stability criterion the flow is hydrodynamically stable for
 $\mu\geq\etah^2$.
 We are only  interested in hydrodynamically
 stable regimes so that  $\mu>\etah^2$ should always be  fulfilled. Rotation laws with ${\rm d}\Om/{\rm d}R>0$  are described by $\mu>1$ while rotation laws with ${\rm d}\Om/{\rm d}R<0$  are described by $\mu<1$. Rigid rotation means $\mu=1$.  Hydrodynamical flows  with rigid rotation or super-rotation are always linearly stable. 
 
 Also the possible magnetic profiles are restricted. The solution of the stationary induction equation without flows reads 
\beg
B_\phi=A  R
\label{basic}
\ende
(in cylinder geometry, see Roberts 1956\cite{R56}, Tayler 1957\cite{T57}) where  
 $A $ corresponds to a uniform axial
current everywhere within $R<R_{\rm out}$.  The quantity
${B_{\rm{out}}}/{B_{\rm{in}}}$
measures the variation of $B_\phi$ across the gap. For  fields after (\ref{basic}) it is simply ${B_{\rm{out}}}={B_{\rm{in}}}/{\rin}$.  
{If the axial electric-current only exists inside the inner cylinder then the solution of the induction equation instead of (\ref{basic}) is 
\beg
B_\phi\propto \frac{1}{R}.
\label{basic2}
\ende
In the present paper  the  stability characteristics   of the MHD system  are   due to the   instability  of the field (\ref{basic}) under the influence of super-rotation.  In order to compare the results for the standard profile (\ref{basic}) with those for the field which is current-free in the fluid, the  profile (\ref{basic2}) has only been used below for data  given in  Fig. \ref{f7}.}
We know that the nonaxisymmetric Tayler instability (TI) also exists for  resting fluids with  threshold values which  do not depend  the magnetic Prandtl number
\beg
\rm Pm=\frac{\nu}{\eta},
\label{pm}
\ende
 the value of which, however, has an essential influence on the excitation of the TI under the influence of rotation \cite{RS10}.  For fast rotation,  a narrow gap, $\Pm\neq 1$ and strong shear we shall present instability maps   which    hardly differ for various magnetic profiles, the vacuum fields of AMRI included. For these solutions, therefore, 
 the importance  of the electric-current inside the fluid disappears and the instability gets its entire energy from the differential rotation. This conclusion will be  supported by the inspection of the associated wave numbers and drift velocities of the various nonaxisymmetric instability patterns.

 Both AMRI and (resting) TI have recently  been realized in the MHD laboratory using the liquid eutectic alloy GaInSn with $\Pm = 1.4 \cdot 10^{-6}$ as the conducting fluid \cite{S12,S14}.
  If the results shall be applied to turbulent media like   the  stellar convection zones then the magnetic Prandtl number must be replaced by its turbulence-induced values which are much larger\cite{YBR03}. In the upper part of the solar core the molecular value  is about $\Pm\simeq 0.065$\cite{G07}.

\section{Equations}
The dimensionless incompressible MHD equations are
 \beg
 {   {\rm Re} \left(   \frac{\partial\vec{u}}{\partial t} +(\vec{u} \cdot \nabla)\vec{u}\right) =
               -\nabla P + \Delta \vec{u} + {\Ha^2}
               \rot \vec{B} \times \vec{B},} \nonumber
\ende
\beg	       
   {\rm Rm} \frac{\partial \vec{B}}{\partial t} =
	        \rot (\vec{u} \times \vec{B}) +\Delta \vec{B}, 
\label{mhd}
\ende
with $\div{\vec{u}} =  \div{\vec{B}} = 0$ and
  with the Hartmann number
 \beg
    \Ha = \frac{B_{\rm in} D}{\sqrt{\mperm \rho \nu \mdiff}}.
    \label{Ha}
\ende
  $D=\sqrt{\Rin(\Rout - \Rin)}$ is used as the unit of length, $\Om_{\rm in}^{-1}$ as the unit of the time, $\eta/D$ as the 
 unit of velocity and $B_{\rm in}$ as the unit of magnetic fields. In this notation the angular velocity   of the global  rotation $\Om$  at the inner cylinder equals $\rm Rm$. The Reynolds number
 $\Rey$ is defined as 
\beg
\Rey=\frac{\Om_{\rm in}  D^2}{ \nu} 
\label{Rey}
\ende
and the magnetic Reynolds number as $\rm Rm=\rm Pm \ \rm Re$. It is also useful to work with the mixed Reynolds number 
\beg
\rm \overline{Rm}=\sqrt{\rm Re Rm}
\label{rue}
\ende
which is symmetric in $\nu$ and $\eta$ as it is  the Hartmann number.  Its ratio to  (\ref{Ha}) is called the magnetic Mach number $\rm Mm$  which measures the rotation rate 
in comparison with the \A\ frequency $\Om_{\rm A}=B_{\rm in}/\sqrt{\mu_0\rho D^2}$,
  \beg
\rm Mm=\frac{\Om_{\rm in}}{\Om_{\rm A}}=\frac{\overline{Rm}}{\Ha}.
\label{Mm}
\ende
We always use no-slip boundary conditions for the velocity 
$
u_R=u_\phi=u_z=0$. The material of the cylinders is assumed as made from  perfect conductors, or in some other cases made from perfect insulators.  
For the conducting walls the fluctuations $\vec{b}$ have thus  to fulfill the conditions 
$
{\rm d} b_\phi/{\rm d}R + b_\phi/R=b_R=0$ 
at both $R_{\rm in}$ and $R_{\rm out}$. 
Mathematical details about the much more complicated vacuum  boundary conditions   and the used numerical  codes can be found in previous publications. The time-tested code for the linearized equations\cite{RKH13} solves the eigenvalue problem for the { Fourier modes  $\exp({\rm i}(\omega t +kz+m\phi))$ where $k$ is the axial wave number and $m$  the azimuthal mode number. 

The nonlinear simulations have been done with our reliable time-stepping code. 
It works with an expansion of the solution in 
Fourier modes in the azimuthal direction generating  a  sample of meridional problems 
each of which is solved using a Legendre spectral element method\cite{F05,GRF07}.} 
 \section{A double-diffusive instability}
\begin{figure}[htb]
     \includegraphics[width=8.0cm]{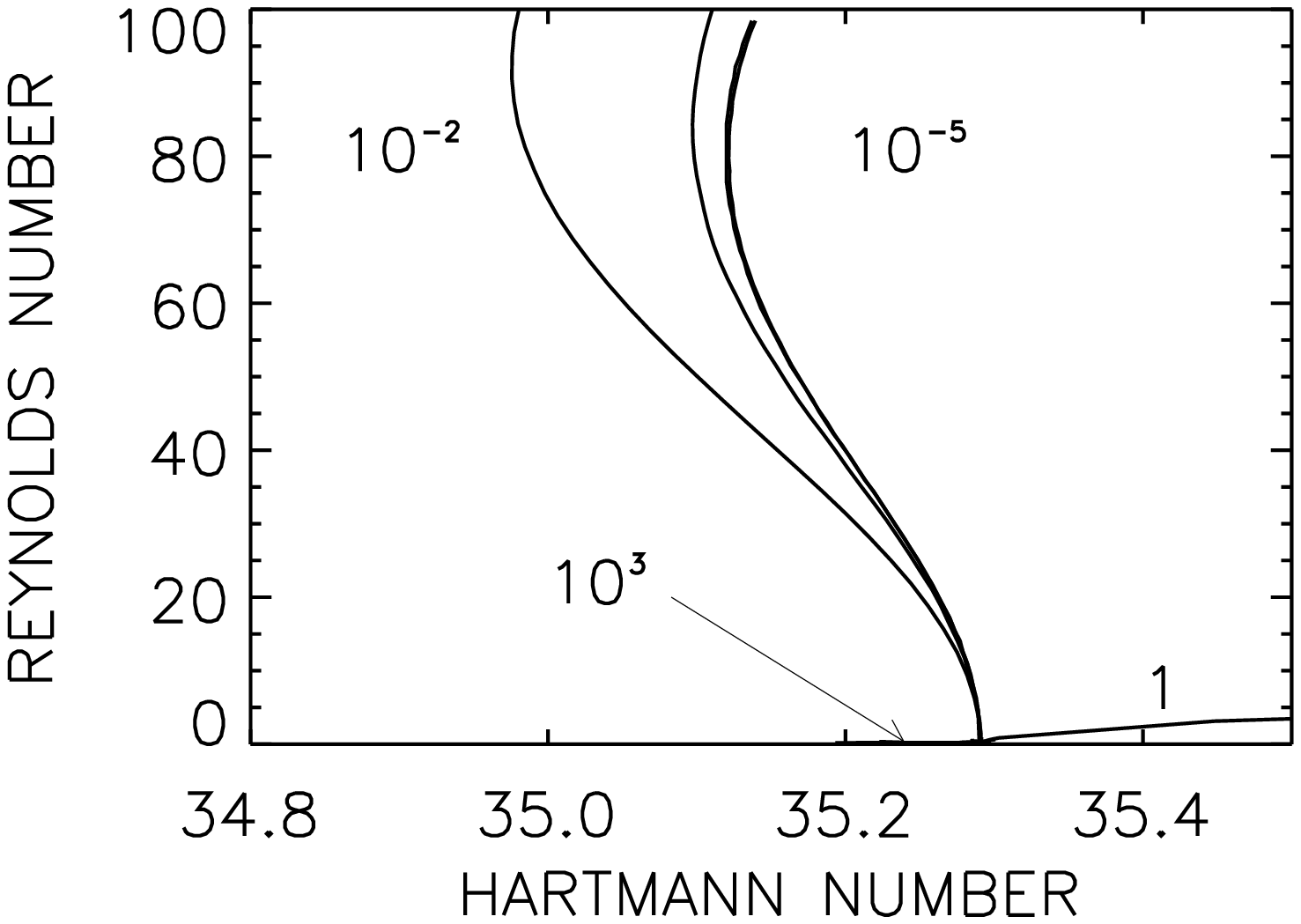}
       \includegraphics[width=8.0cm]{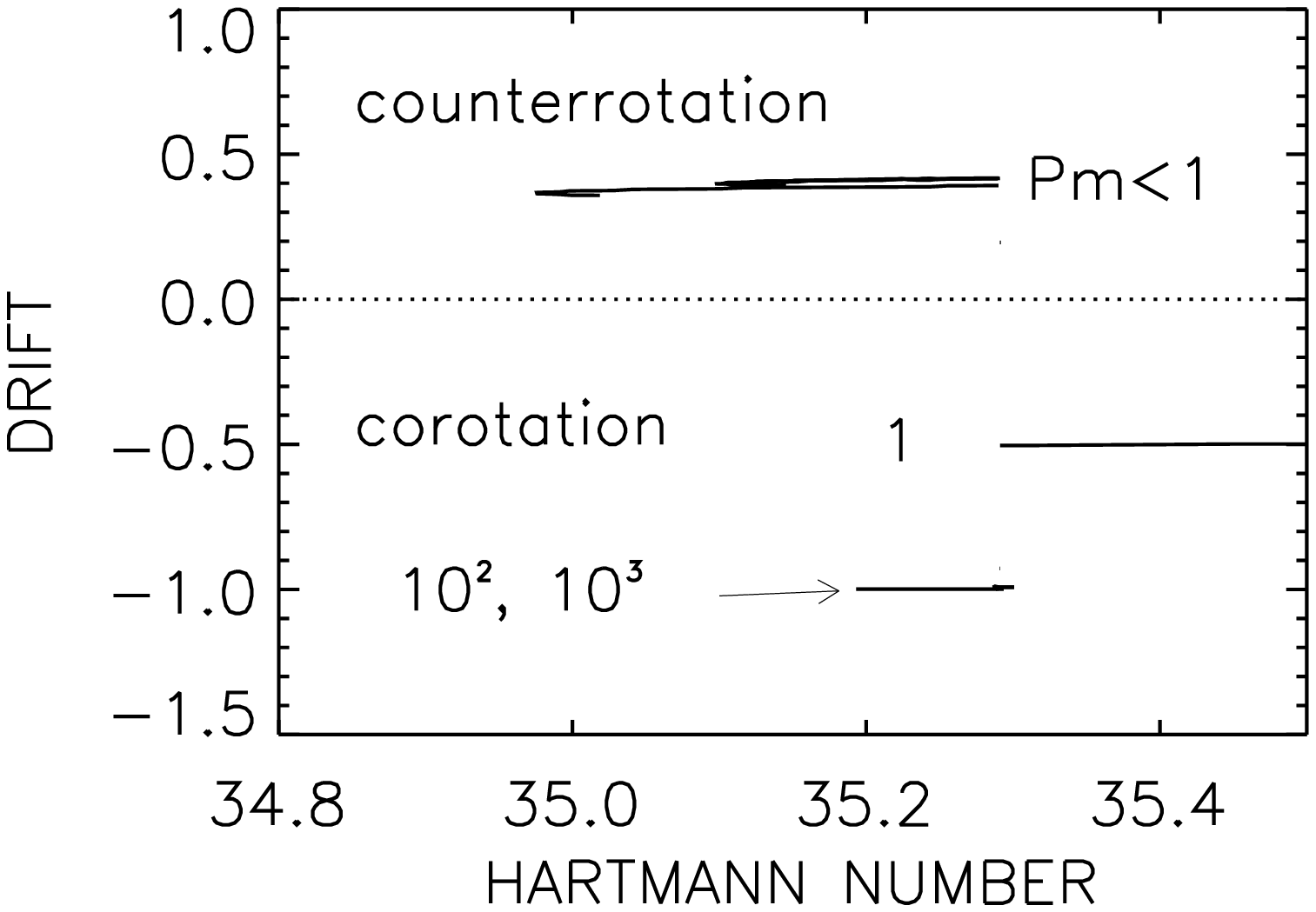}
     \caption{Instability  of the $m=\pm 1$ mode for   rigid rotation. The numbers  represent the magnetic Prandtl number $\Pm$. Left panel: The lowering of the critical Hartmann numbers by the global rotation for $\Pm\neq 1$ Note the maximal relaxation  of the $\Ha$ existing  for $\Pm=O(10^{-2})$ but not for  $\Pm\to 0$. Right panel: The drift rates $\omega_{\rm dr}/\Om$ of the modes for marginal instability for small $\Pm$ and large $\Pm$. The  axial wave numbers $k$ for all points at the lines do hardly vary. $\rin=0.5$,  perfect-conducting boundaries.}
    \label{rigid}
 \end{figure}

 We start with solid-body rotation of a container. It is known that this flow with (\ref{basic}) belongs to the  class of magnetized flows where the   radial profiles of
 the global velocity and the global magnetic field are identical. Chandrasekhar (1956) has shown that all {\em flows} of this class are stable\cite{C56}. One can also show for the nonideal  MHD flows that  all   lines of marginal instability for very small $\Pm$ are identical in the $\Ha$-$\Rey$ plane\cite{R15}. They are given for various magnetic Prandtl numbers in Fig. \ref{rigid} (left panel).  For small $\Pm$ the lines for $\Pm<10^{-3}$  cannot be separated optically. It is also demonstrated that the critical Hartmann number $\Ha_{\rm Tay}$ for $\Rey=0$ does not depend on $\Pm$.  The curves, however, for $\Rey>0$ behave different. For $\Pm<1$  they turn to the left while for $\Pm\geq 1$ they are turning to the right. In the former case the instability is supported by the rotation and in the latter  case it is (strongly) suppressed. Note that the lowering of the Hartmann number by global rotation only exists for $\Pm\neq 1$ and for slow rotation with $ \Mm\ll 1$. For those parameters ``the stabilizing effect of global rotation is greatly reduced", as it already has been formulated for this kind of double-diffusive problems\cite{AG78}. A very similar behavior also appeared for the excitation of axisymmetric modes for helical background fields ($B_z B_\phi \neq 0$) under the influence of differential rotation which are also stable for $\nu=\eta=0$ and their critical eigenvalues are lowered for $\Pm\neq 1$  compared with those for  equal diffusivities\cite{LV07}.
 
 For faster rotation all curves are turning to the right so that the statement finally becomes   correct that generally the rotation suppresses the Tayler instability.

 Interesting is also the behavior of  the  drift rate $\omega_{\rm dr}$ as the real  part of the frequency $\omega$ of the Fourier mode of the instability pattern,  normalized with the rotation rate of the   cylinders.  It can be  positive  or negative. Because of the definition 
\beg 
 \dot \phi=-\frac{\omega_{\rm dr} \Om_{\rm out}}{m}
 \label{drift}
 \ende 
the azimuthal migration of the instability pattern has the opposite sign of  $\omega_{\rm dr}$. The solutions for the modes with $m=1$ and $m=-1$ have the same eigenvalues $\Rey$ and $\Ha$ but the opposite signs of $\omega_{\rm dr}$. After (\ref{drift}) they have thus the same frequency of migration in $\phi$-direction. Figure  \ref{rigid} (right panel) shows that the drift rate hardly depends on the Hartmann number but it is strongly directed by the magnetic Prandtl number $\Pm$.  It is positive for small $\Pm$ which with (\ref{drift}) leads to an azimuthal migration of the instability pattern opposite to the cylinder  rotation  but for $\Pm \geq 1$ it changes the sign so that the instability pattern rotates in the same direction as the cylinders do. The relaxation of the critical Hartmann number for $\Pm\neq 1$ is { not} indicated by the drift rates.
 
\begin{figure}[htb]
     \includegraphics[width=8.0cm]{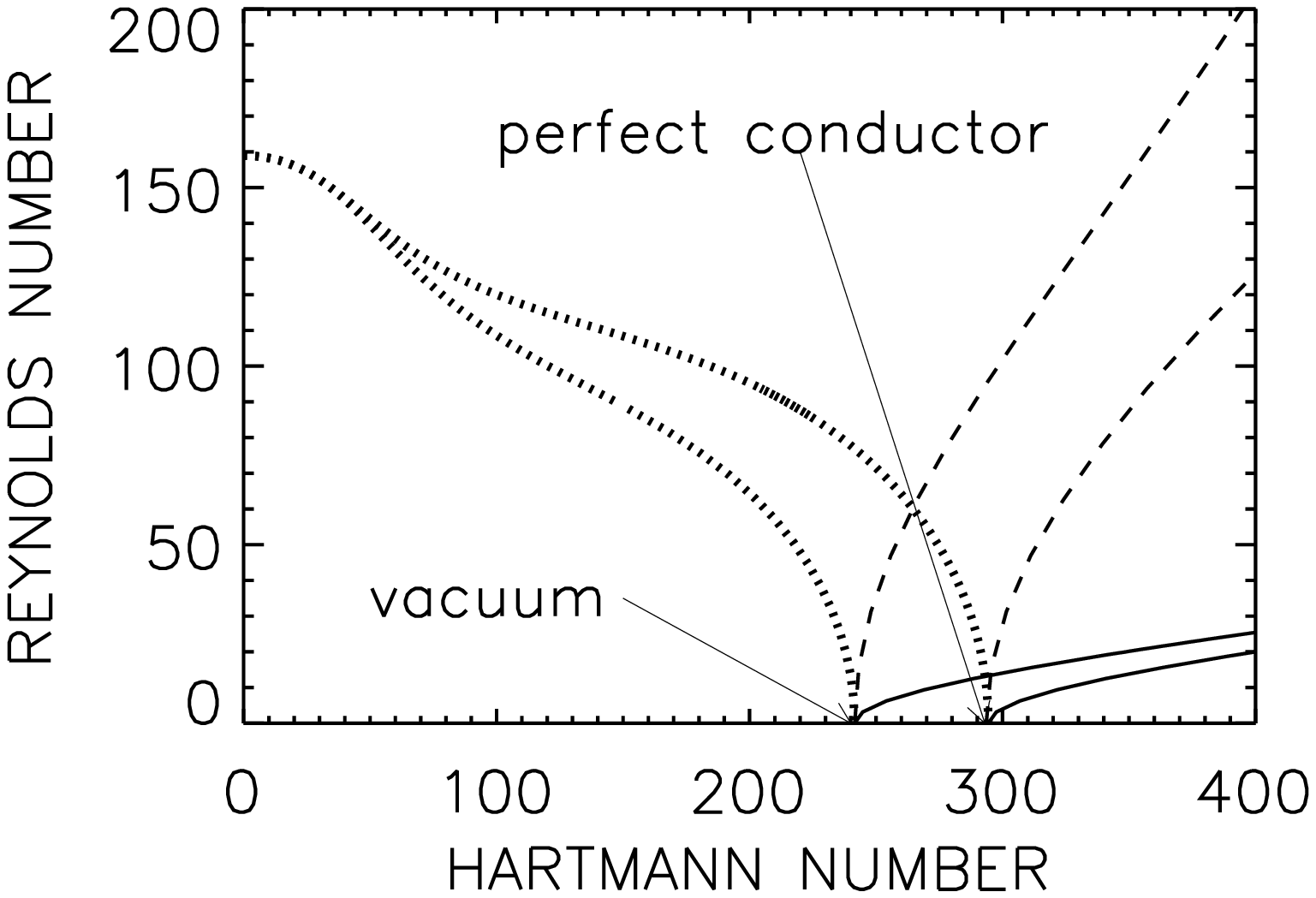}
       \includegraphics[width=8.0cm]{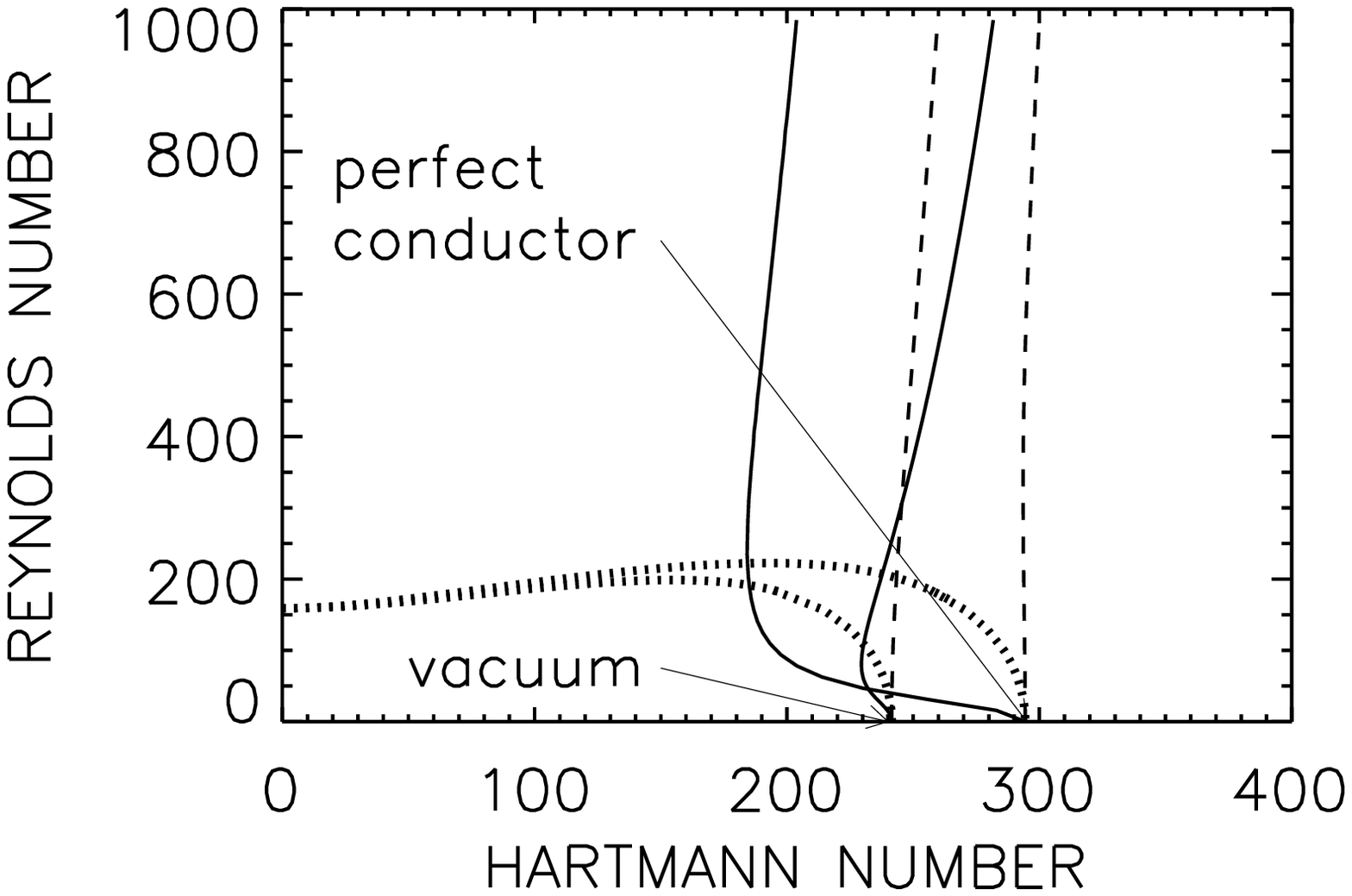}
     \caption{Instability map for the $m=\pm 1$ mode for   $\rm Pm=1$ (left) and $\rm Pm=10^{-5}$ (right).  There are examples  for  sub-rotation ($\mu=0.5$, dotted lines), rigid rotation ($\mu=1$, dashed lines) and super-rotation ($\mu=4$, solid lines). The lowering of the critical Hartmann numbers for super-rotation only appears for $\Pm\neq 1$. The  boundary conditions are those for vacuum or for  perfect conductors, $\rin=0.8$.}
    \label{f9}
 \end{figure}

Figure \ref{f9} summarizes  the influence of  {\em nonuniform  rotation} on the excitation of the TI against nonaxisymmetric modes with $m=1$. It is $\rin=0.8$ and  the cylinders are made from insulating  or perfect-conducting materials. The critical Hartmann number for resting cylinders is $\Ha_{\rm Tay}=250$ for vacuum boundary conditions and $\Ha_{\rm Tay}=290$ for perfect-conductor conditions. Again, these values  do not depend on the magnetic Prandtl number   as here demonstrated here for the two examples with $\Pm=1$ (left panel) and $ \Pm=10^{-5}$ (right panel).  One finds that the TI  can be excited more easily for vacuum boundary conditions than for perfect-conducting cylinders.

 The dotted lines in Fig. \ref{f9} are the lines of marginal instability  for the sub-rotation law with $\mu=0.5$.  In the narrow gap and for fast enough rotation such a  rotation law is  linearly unstable  without magnetic field. For not too small $\Pm$ the magnetic  field even destabilizes  such a steep sub-rotation law so that for finite Hartmann number the critical Reynolds number is always lower than 160 which is the critical value for $\Ha=0$. The influences of the boundary conditions and the magnetic Prandtl number on the stability/instability  of sub-rotation laws is only small.

The solid lines are due to  rotation laws with positive radial shear.  Their behavior {\em strongly} depends on the value of the magnetic Prandtl number. For $\Pm=1$ (left panel) super-rotation ($\mu=4$) acts stabilizing (${\rm d} \Rey/{\rm d} \Ha>0$ everywhere) while for small magnetic Prandtl number ($\Pm= 10^{-5}$, right panel) it acts destabilizing (${\rm d} \Rey/{\rm d} \Ha<0$  in the lower part of the diagram).  Under the presence of rotation with positive shear the electric-current  becomes unstable for lower Hartmann numbers than for the resting pinch. This `subcritical excitation´  of the TI for super-rotation is insofar interesting as TC-flows with positive shear are prominent examples of stable hydrodynamic flows\cite{SG59}. Such a very stable configuration can even  be destabilized by a magnetic field  which is weaker than the critical field  for the TI. This phenomenon only exists for slow rotation since  for fast rotation all nonaxisymmetric magnetic instabilities are suppressed by any sort of differential rotation. The magnetic Mach number $\rm Mm$ for the subcritical excitation by super-rotation in the right panel of Fig. \ref{f9} is (only) of order $10^{-3}$. Note that a rotation profile with $\mu>1$  needs { much higher Hartmann numbers   to be destabilized than a rotation law with $\mu<1$ and sufficiently high enough Reynolds number}. With other words, the toroidal field which is induced by a super-rotation  can become much stronger than the toroidal field which is induced by a sub-rotating $\Om$-profile. This basic finding should have implications for the electrodynamics of rotating stars. With other words, the instability {\em requires} rotation laws with negative shear in order to exist for large magnetic Mach numbers.

The dashed lines describe the influence of rigid rotation which for fast rotation  always acts stabilizing, i.e. ${\rm d} \Rey/{\rm d} \Ha>0$. This effect, however,  is much weaker for small $\Pm$  than for $\Pm=1$. In the latter case the instability is suppressed  for all  slow rotation rates.  The stabilization depends on the value 
of the magnetic Prandtl number,  it is strongest for $\rm Pm=1$.

In summary,  we have found that  for $\Pm=1$ rigid rotation and super-rotation suppresses the TI for all $ \Rey$. Rigid rotation and super-rotation support  the TI for $\Pm\neq 1$ but only for slow  rotation by lowering the Hartmann number,  $\Ha<\Ha_{\rm Tay}$, below the value which holds for the resting pinch.  The lowering only exists if the two molecular diffusivities have different values -- no matter which is larger or smaller --   typical for a double diffusive instability\cite{AG78,S09}. 

From Fig. \ref{f9}  one also finds that the boundary conditions do not play a minor role. For perfect-conducting cylinders the lowering of the Hartmann numbers is much stronger than than for cylinders of insulating material.  
In order to study the influence of the geometry on the instability  we shall consider in more detail the two  TC-flows with a rather  narrow gap ($\rin=0.95$) and with a wide  gap  ($\rin=0.5$) . 

 \section{Narrow gap}
\begin{figure}[htb]
     \includegraphics[width=5.4cm]{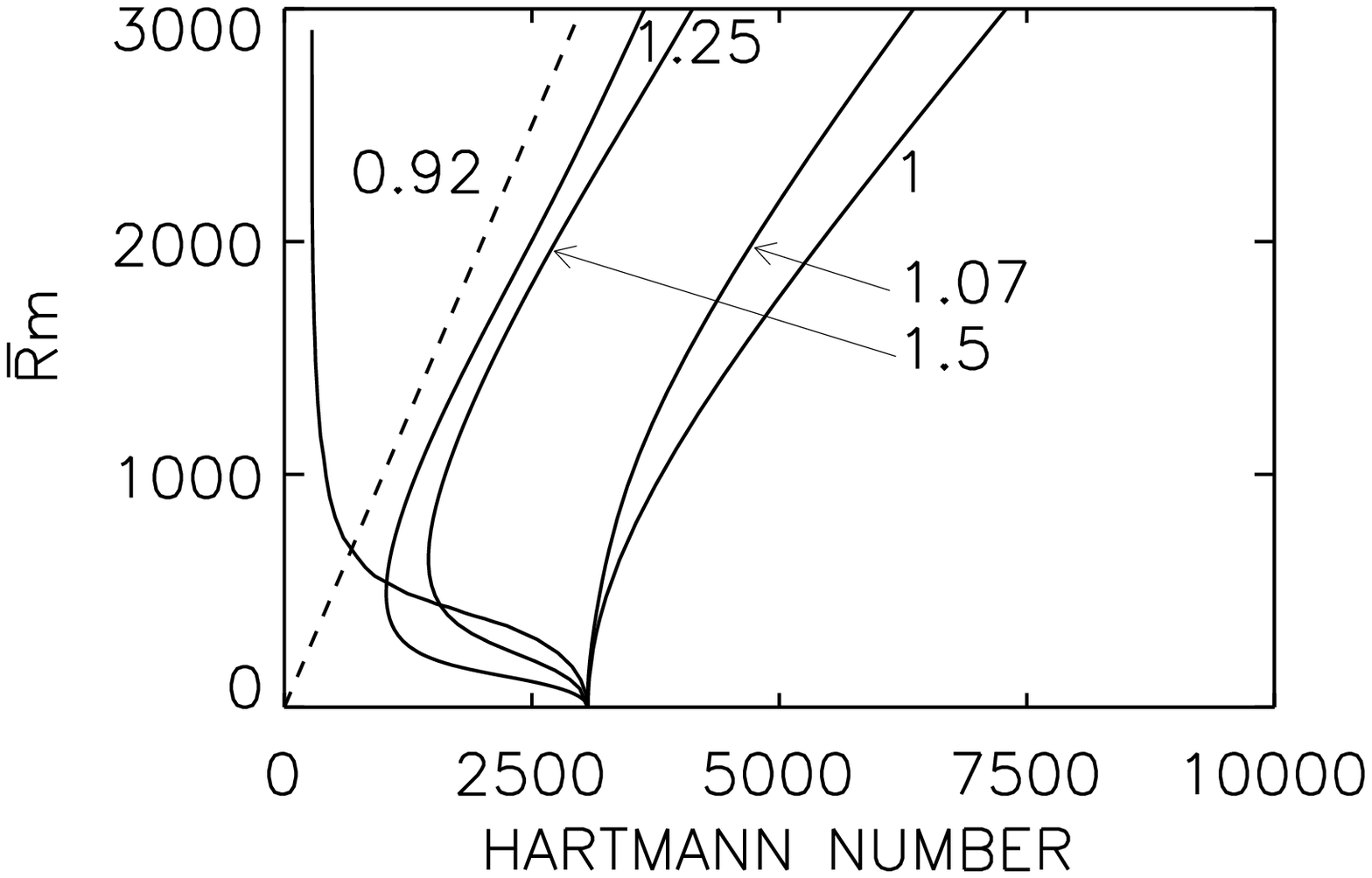}
       \includegraphics[width=5.4cm]{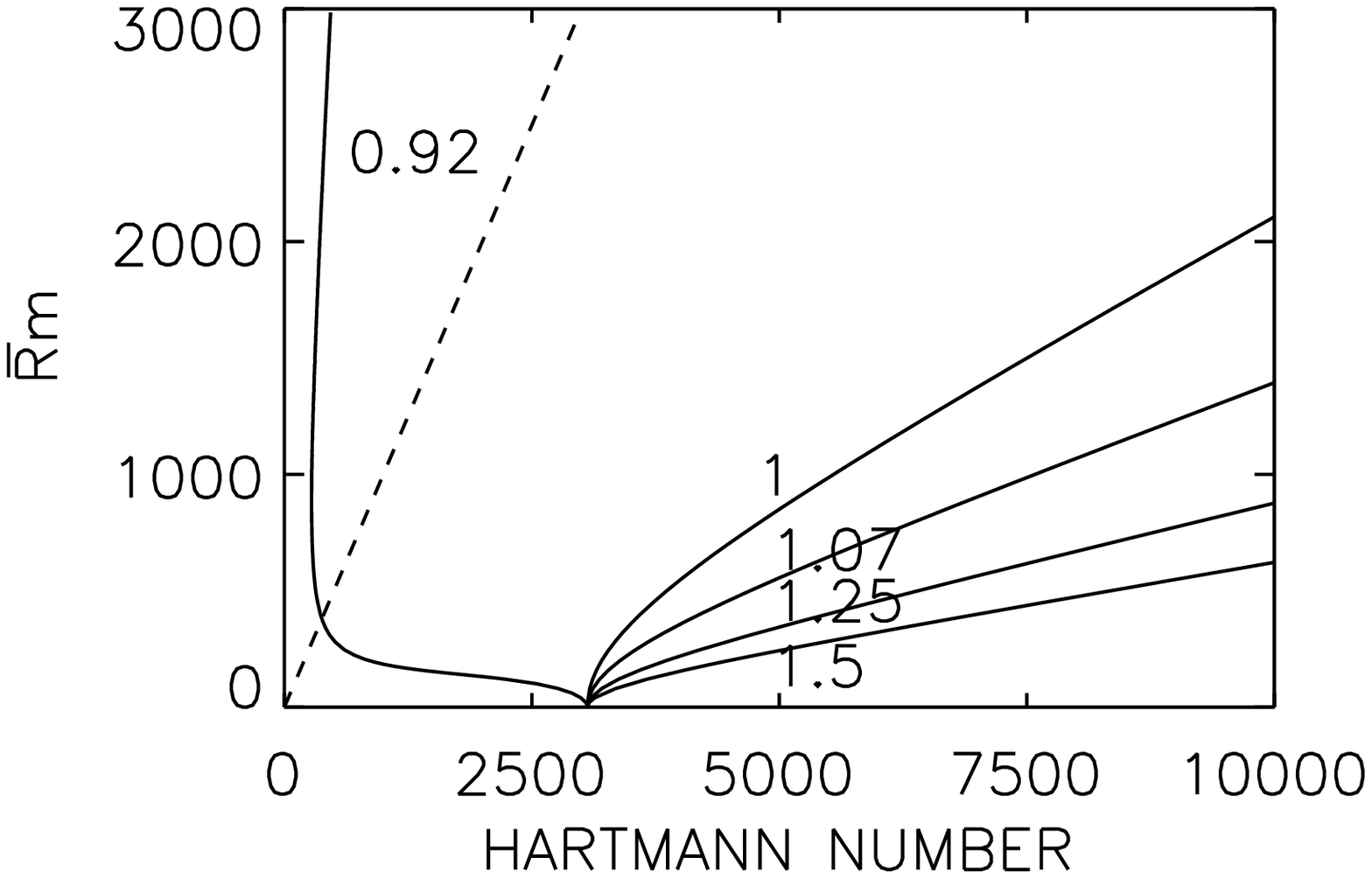}
       \includegraphics[width=5.4cm]{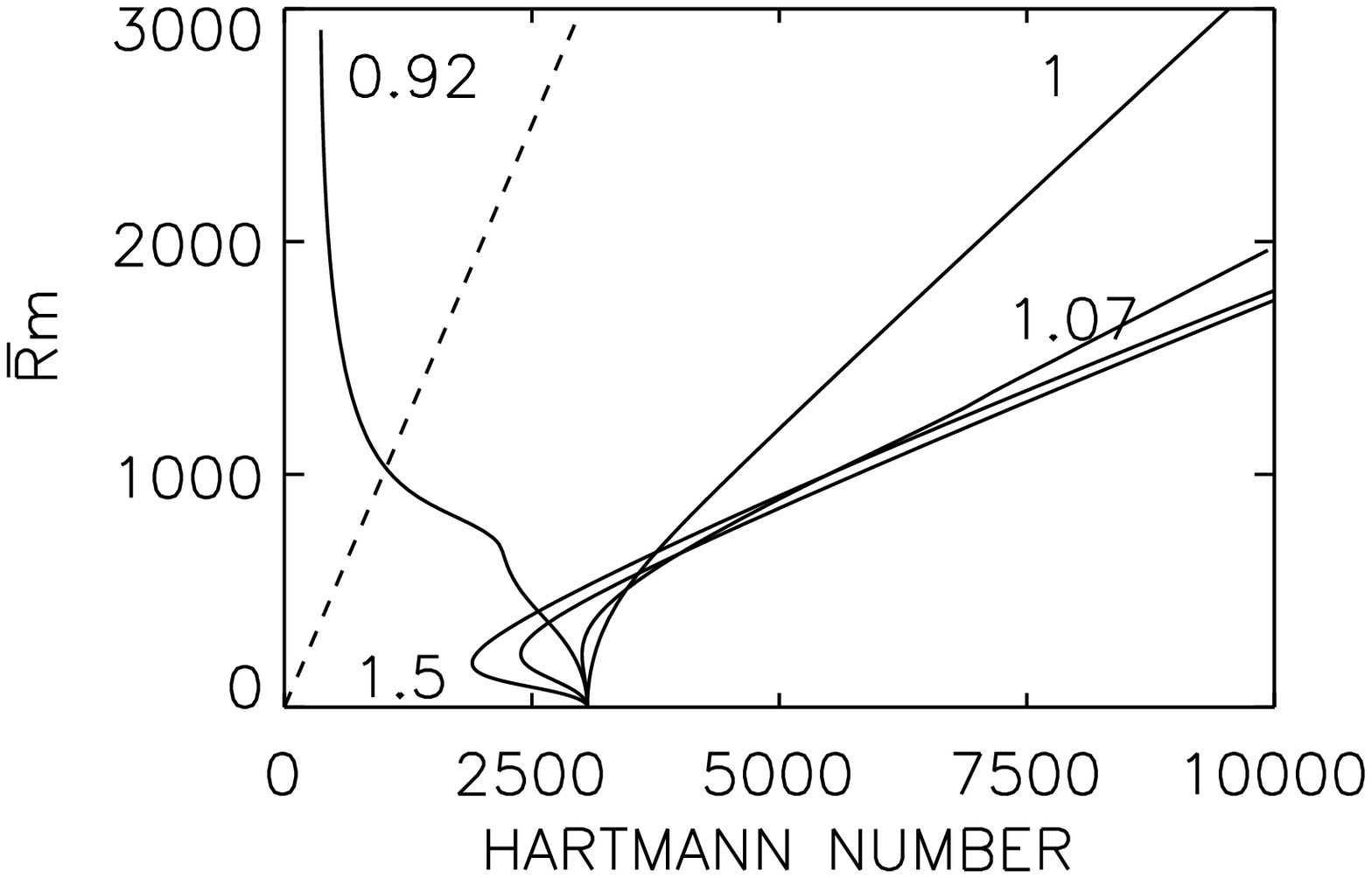}
     \caption{Stabilization and destabilization by super-rotation in a narrow gap  for  $\rm Pm=10$ (left),  $\rm Pm=1$ (middle) and  $\rm Pm=0.1$ (right). The modified Reynolds numbers (\ref{rue}) are used for an easy definition of the unity magnetic Mach number $\Mm$  (dashed lines). For comparison one example for sub-rotation is given which itself is hydrodynamically stable. The curves are marked with their value of $\mu$. $\rin=0.95$, perfect-conducting boundaries. $m=\pm 1$.}
    \label{f8}
 \end{figure}

For a narrow gap the influence of the magnetic Prandtl number on the Tayler instability in a  container with various rotation profiles shall be studied. All the considered rotation laws are hydrodynamically stable.
  For a  gap with $\rin=0.95$  Fig.~\ref{f8} gives the results for $\rm Pm=0.1$, $\Pm=1$ and $\rm Pm=10$. The critical Hartmann number for TI without rotation is $\Ha_{\rm Tay}=3060$. For $\Pm=1$ rigid-body rotation and super-rotation of any Reynolds number are always stabilizing, i.e. $\Ha>3060$. Only sub-rotation leads to  $\Ha<3060$. The differences for both  sub-rotation and for super-rotation here only appear for rather low Reynolds numbers.
  
   It is also worth to mention that the lines of marginal instability for rigid rotation and for super-rotation { always lie  {\em below} the line $\Ha=\overline{\Rm}$}, i.e. even slow rotation stabilizes the TI for $\Pm=1$. For fast rotation the TI only exists under the presence of differential rotation with negative shear.

\subsection{Subcritical excitations}
To discuss the results for  $\Pm\neq 1$ it  makes sense  to use the modified Reynolds number $\overline{\rm Rm}$ for the characterization of the basic rotation.  For both $\Pm> 1$ (Fig. \ref{f8}, left panel) and $\Pm< 1$ (Fig. \ref{f8}, right panel)  also the  rotation laws with positive shear lead to subcritical excitations if the rotation is slow enough. The magnetic Mach number which measures the rotation of the (inner) cylinder to the \A\ frequency for the subcritical excitation and   for both  magnetic Prandtl numbers is $\rm Mm\simeq 0.05$. Again the curves for rigid rotation and for super-rotation are located below the line $\Mm=1$. Again, for $\Mm>1$ the TI needs the action of a sub-rotation law with negative shear.

For sufficiently  fast rotation  the super-rotation laws are always stabilizing. The super-rotation for   small magnetic Prandtl numbers is  much more  stabilizing than that for high magnetic Prandtl numbers. For $\rm Pm=10$ the stabilization by super-rotation is even weaker than that of rigid rotation. It is often the rule  for magnetic instabilities that large  $\Pm$ destabilize nonuniform rotation while small  $\Pm$ stabilize the flows.  The formal reason for this phenomenon can be realized in Fig.~\ref{f9} where the bifurcation curve for super-rotation for small $\Pm$ moves to the left of the line for rigid rotation rather than to the right as for $\Pm=1$.

It is known that  rotation laws  with negative shear (here $\mu=0.92$) behave strongly  destabilizing the flow. The domain of stability in Fig.~\ref{f8} is again larger for small $\rm Pm$.  For sufficiently fast rotation also the lines of marginal instability for sub-rotation turn to the right stabilizing the system as rotation laws of strong shear of both signs do always erode nonaxisymmetric magnetic patterns.

\begin{figure}[htb]
     \includegraphics[width=8.0cm]{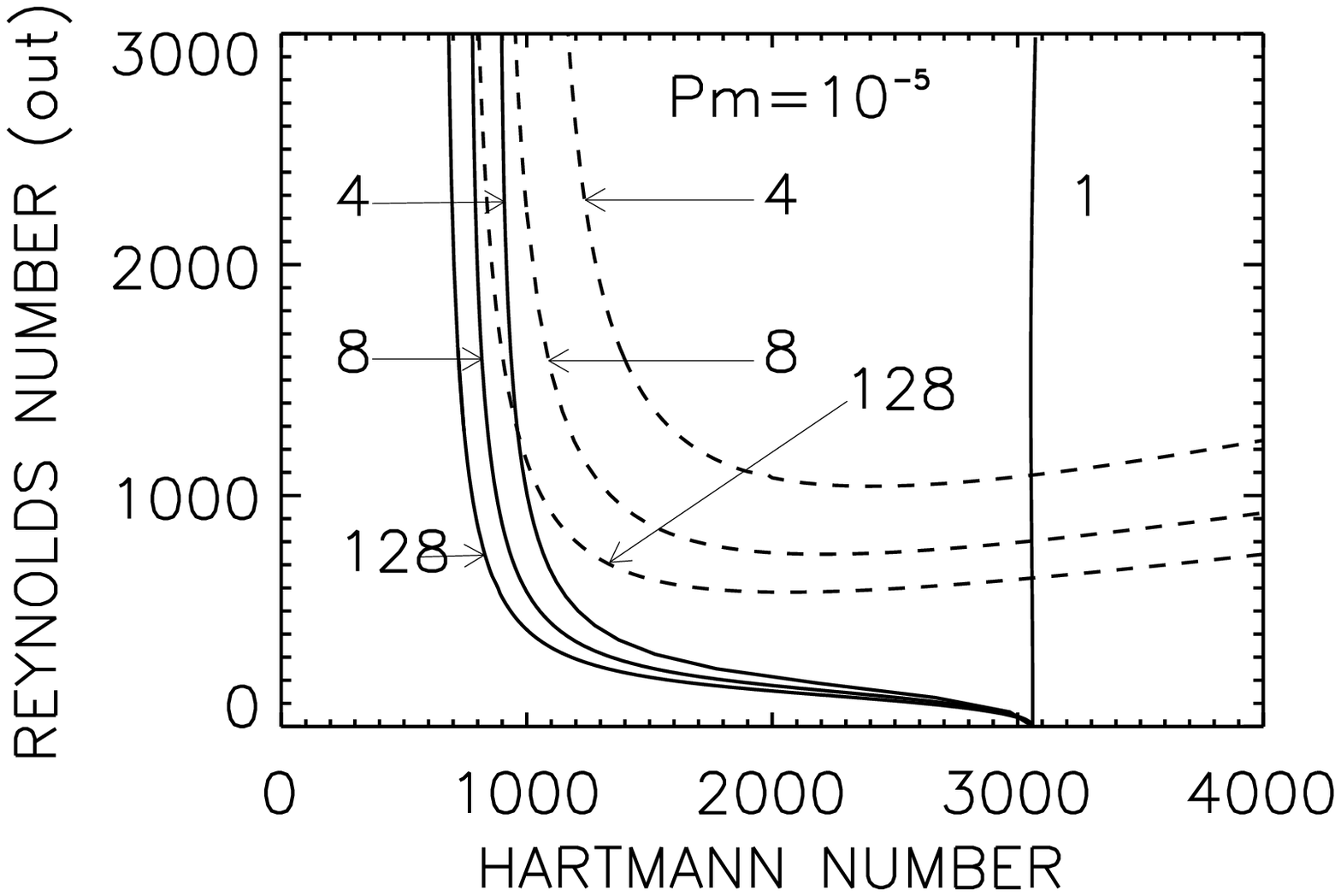}
          \includegraphics[width=8.0cm]{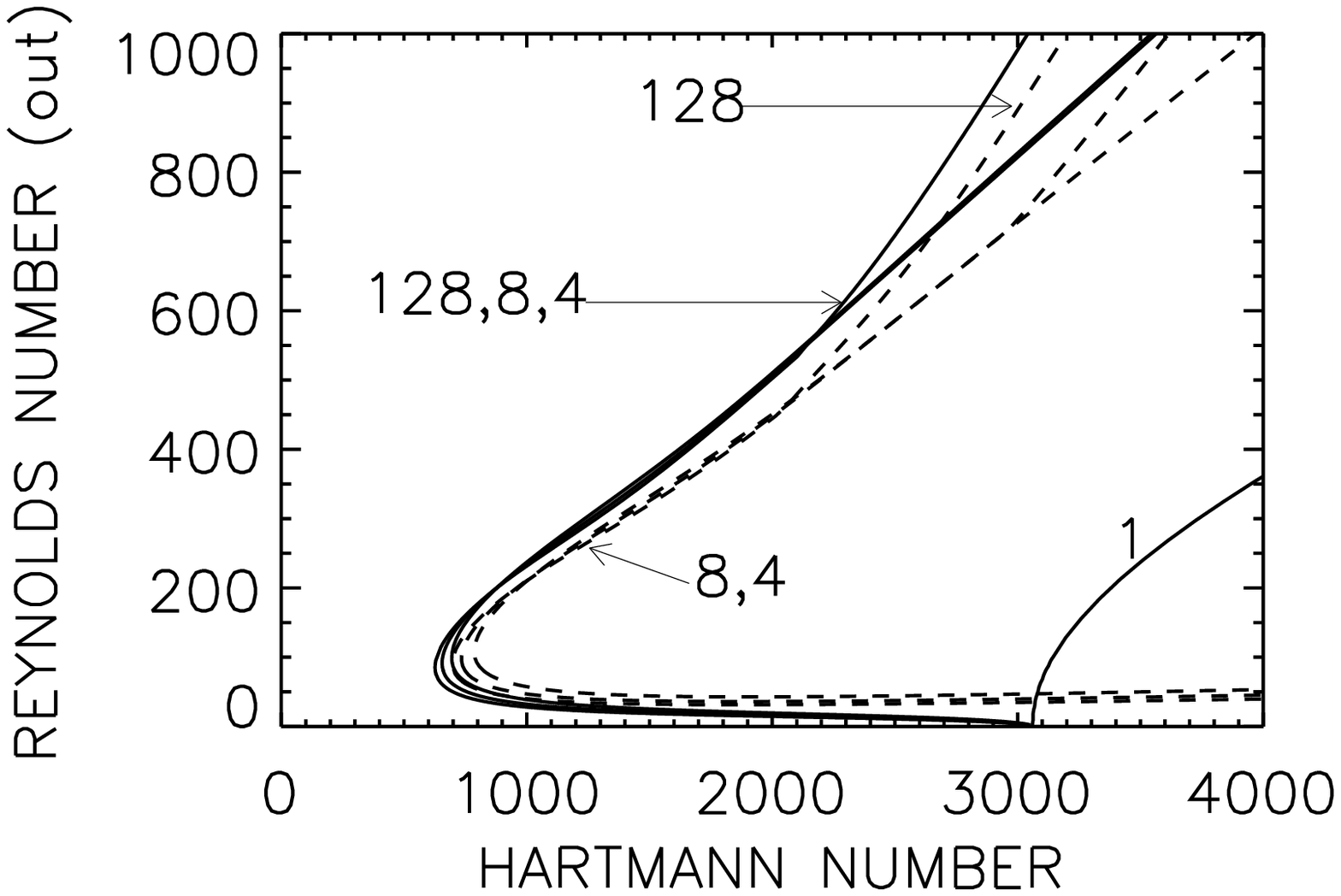}
     \caption{Instability  maps  for   super-rotation for small $\Pm$ ($\rm Pm=10^{-5}$, left) and large $\Pm$ ($\Pm=10$, right), the  lines  marked with their values of  $\mu$. The curves are given for the  two  extreme  radial profiles  (\ref{basic})  and  (\ref{basic2}) by solid and by dashed lines.  Here, the Reynolds numbers are defined with   the {\em  outer} rotation rate. $\rin=0.95$, perfect-conducting cylinders.}
    \label{f7}
 \end{figure}

The  question arises about the possible existence of a minimum Hartmann number for steeper and steeper super-rotation  laws. The existence of such a limit is suggested by  the suppression of a nonaxisymmetric magnetic field by differential rotation which should  grow with growing values of shear. The line of marginal instability can never touch the vertical axis as without magnetic field  super-rotation always behaves stable. Figure \ref{f7}  shows very close lines for  $\mu=4$,   $\mu=8$ and even  $\mu=128$ so that  the minimum Hartmann number $\Ha_{\rm Min}$ can be estimated   with strongest super-rotation as smaller by a factor of three compared with  $\Ha_{\rm Tay}=3060$. For the  very small magnetic Prandtl number used for Fig. \ref{f7} (left) the numerical value of $\Ha_{\rm Min}/\Ha_{\rm Tay}$ is astonishing small.  For large $\Pm$  (right panel of Fig. \ref{f7}, $\Pm=10$) the subcritical excitation also occurs with  $\Ha_{\rm Min}/\Ha_{\rm Tay}$  even smaller.  For larger Reynolds number almost all curves (except the curve for rigid rotation) are identical, they  depend on  numerical values of shear and current only in a very little manner. Compared with the curves for very small $\Pm$, however, the curves have a different form. 
 
 There is another striking feature plotted in Fig. \ref{f7}. In the domain where the lines are almost vertical the dependence of the critical Reynolds number on the critical Hartmann number is extremely weak. It is shown in this plot  that there even the dependence of the curves on the radial  profile of $B=B_\phi(R)$ is weak. The dashed lines in    Fig. \ref{f7} represent the instability for the field (\ref{basic2}) which  is current-free between the cylinders within the fluid. These curves, therefore, can never  cross the horizontal axis where $\Rey=0$. For fast rotation, however, they almost coincidence with the solid lines for the marginal instability of the flow with  axial current. Surprisingly, for strong shear and fast rotation the presence of the electric-current  becomes  irrelevant for the occurrence  of   instability. One can show that all possible radial profiles of $B_\phi$ between (\ref{basic}) and (\ref{basic2})  provide more or less the same instability curves in this domain of the bifurcation map revealing  that the differential rotation for $\Pm\neq 1$ is able to deliver the entire  energy for the maintenance  of the instability patterns and the magnetic field only acts as a catalyst. This phenomenon is already known from AMRI for sub-rotation but here, for super-rotation, it only works for $\nu\neq \eta$.
 
The close relatedness of both  the instabilities for the lowest Hartmann numbers  is obvious.  
  It is not yet clear whether the  coincidence of the  lines with and without electric-current in the fluid occurs only for the considered model with a very narrow gap and perfectly-conducting cylinders or not. Note that in narrow gaps the radial profiles of the azimuthal fields  between the cylinders both are almost uniform. Indeed, test calculations also provided instability even for fields with uniform $B_\phi$ in the same domain of Reynolds number and Hartmann number.  One could believe that for $\Pm\neq 1$ the super-rotation becomes unstable under the mere presence of any toroidal field but for $\Pm=1$  the dissipation processes prevent the excitation of this rather slow  (see below) instability. 
 \subsection{Pattern migration}
\begin{figure}[htb]
       \includegraphics[width=8.0cm]{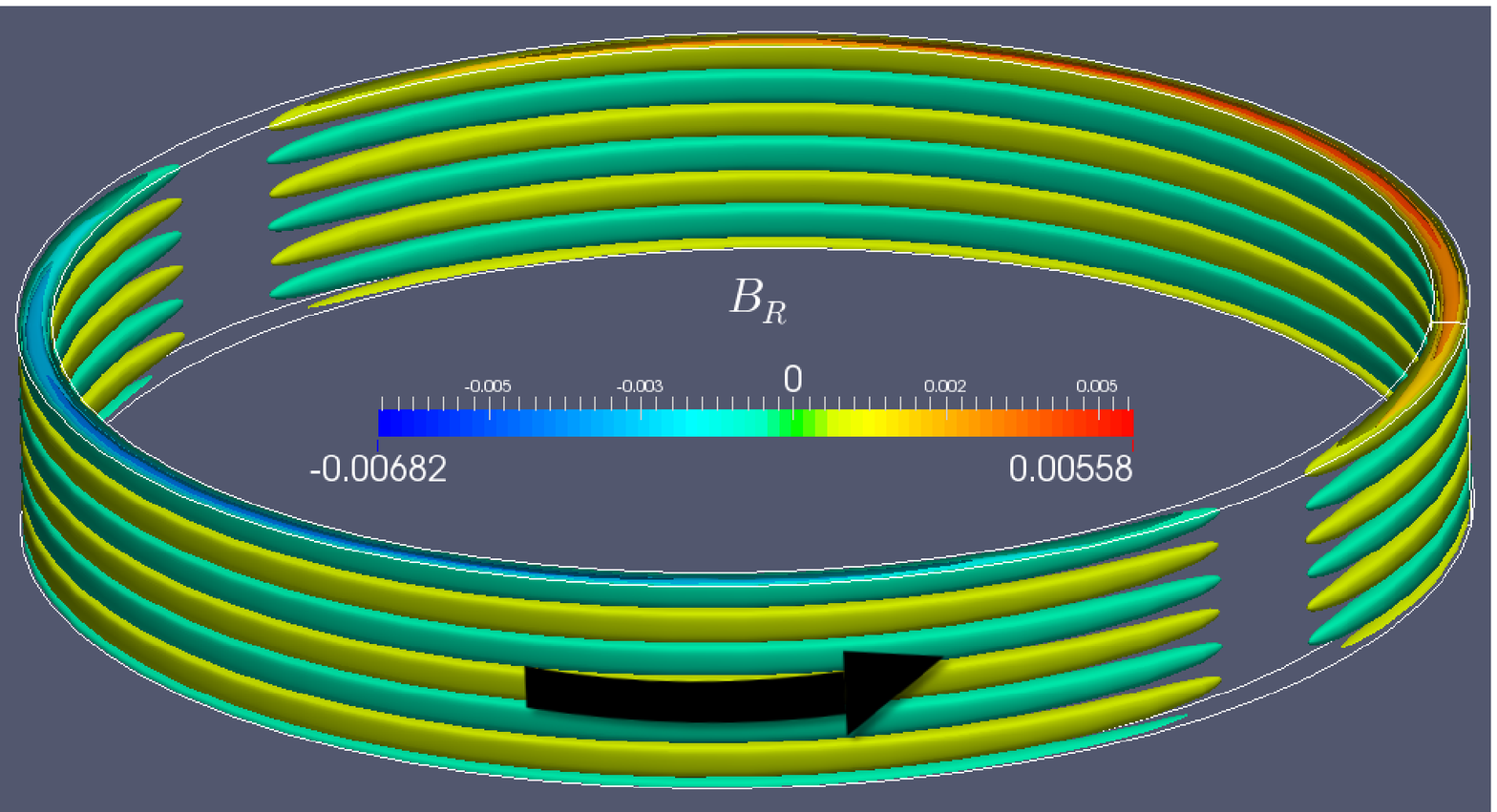}
       \includegraphics[width=8.0cm]{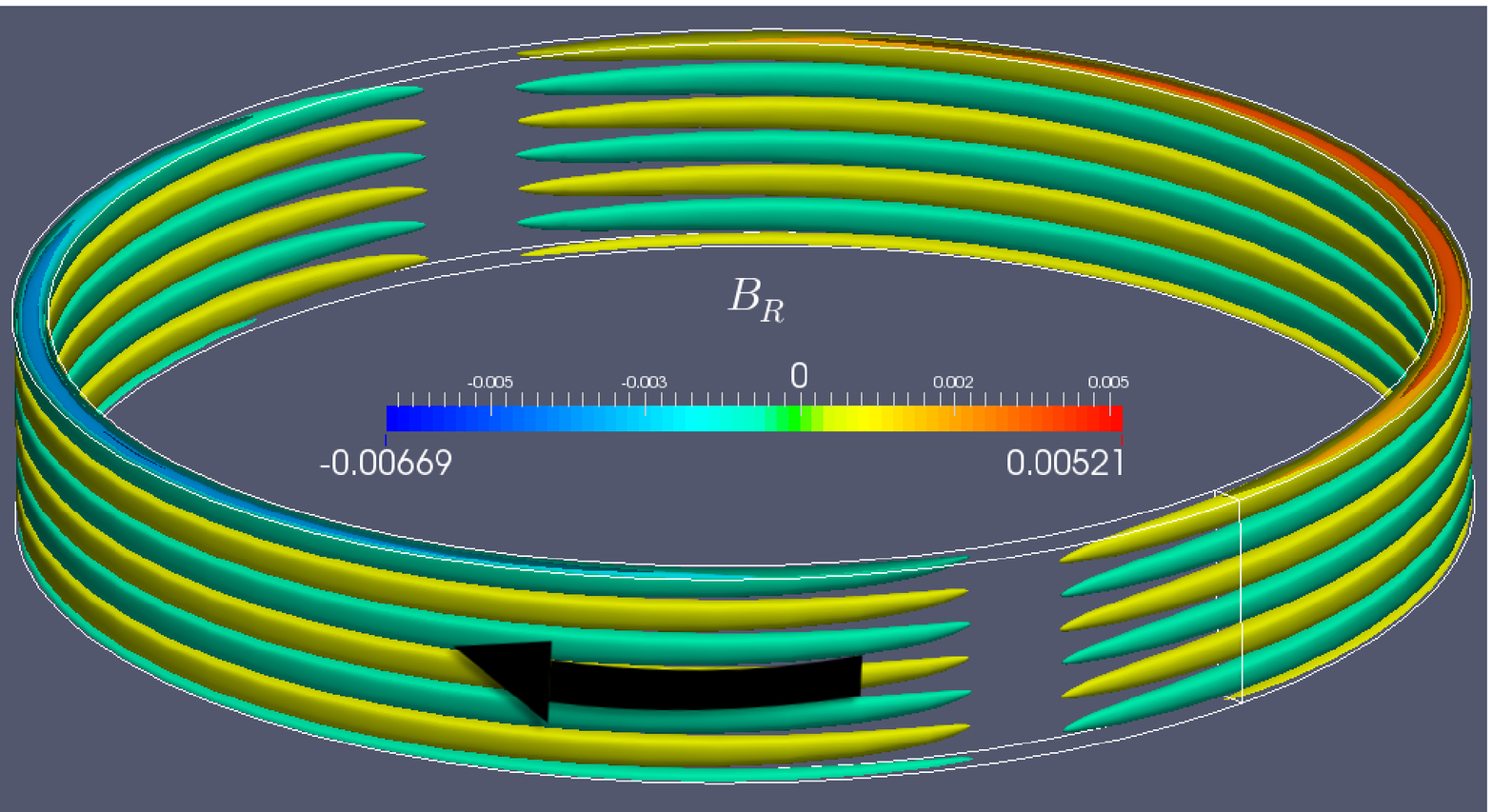}
     \caption{The isolines of the $b_R$ in a narrow gap for  $\rm Pm=1$ (left) and  $\rm Pm=0.1$ (right) for rigid rotation. Both modes with $m=\pm 1$  are excited with the same amplitude forming a standing wave. The pattern with $\rm Pm=1$ migrates in the rotation  direction  while it migrates opposite for $\rm Pm=0.1$ . $\rin=0.95$, $\mu=1$, $\Rey=111$, $\Ha=3440$, perfect-conducting boundaries. }
    \label{f99}
 \end{figure}

 In general both modes with $m=\pm 1$ are simultaneously excited.  If the pinch rotates rigidly then both modes have exactly the same amplitudes and form a standing wave. The instability pattern looks azimuthally dipolar with the drift direction depending on the magnetic Prandtl number. Note also the nearly circular geometry of the resulting cells in the $R$-$z$-plane (Fig. \ref{f99}). For a certain $\Pm$ between 0.1 and 1 the azimuthal migration disappears and the entire pattern  will rest in the laboratory. For nonuniform rotation laws with finite shear one of the modes $m=1$ or $m=-1$ is preferred and the instability pattern  approaches  a spiral.

\begin{figure}[htb]
     \includegraphics[width=5.4cm]{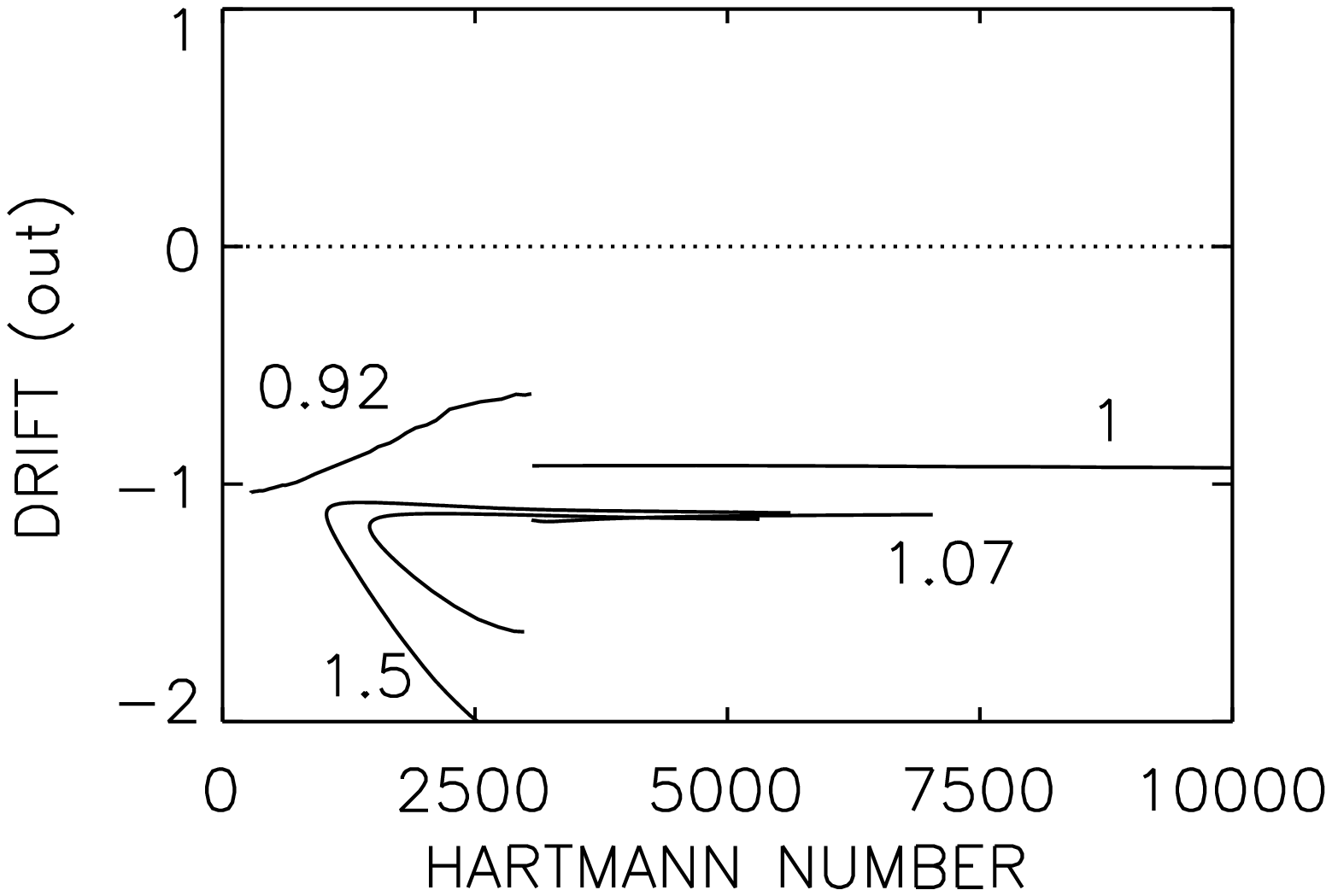}
       \includegraphics[width=5.4cm]{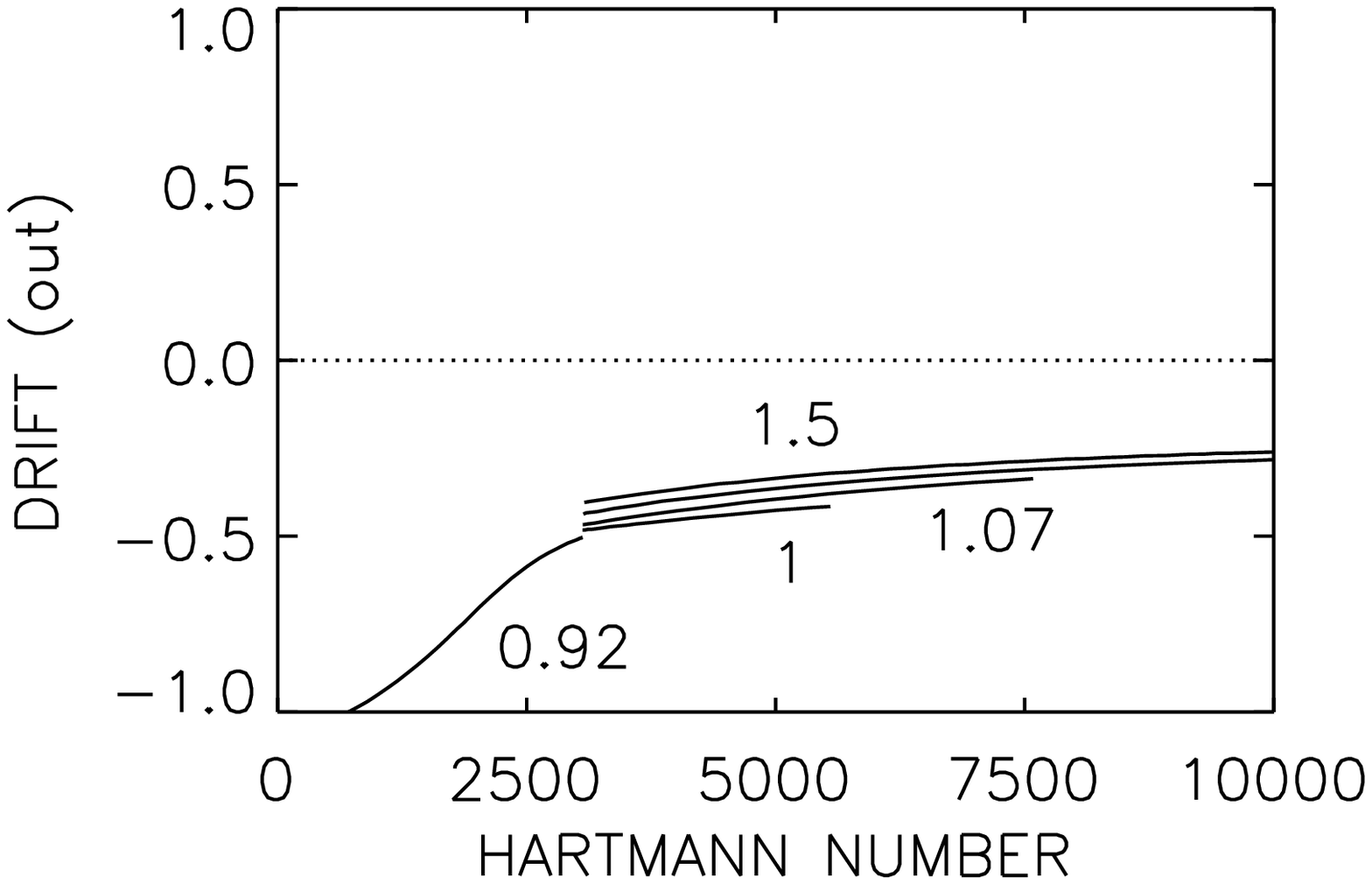}
       \includegraphics[width=5.4cm]{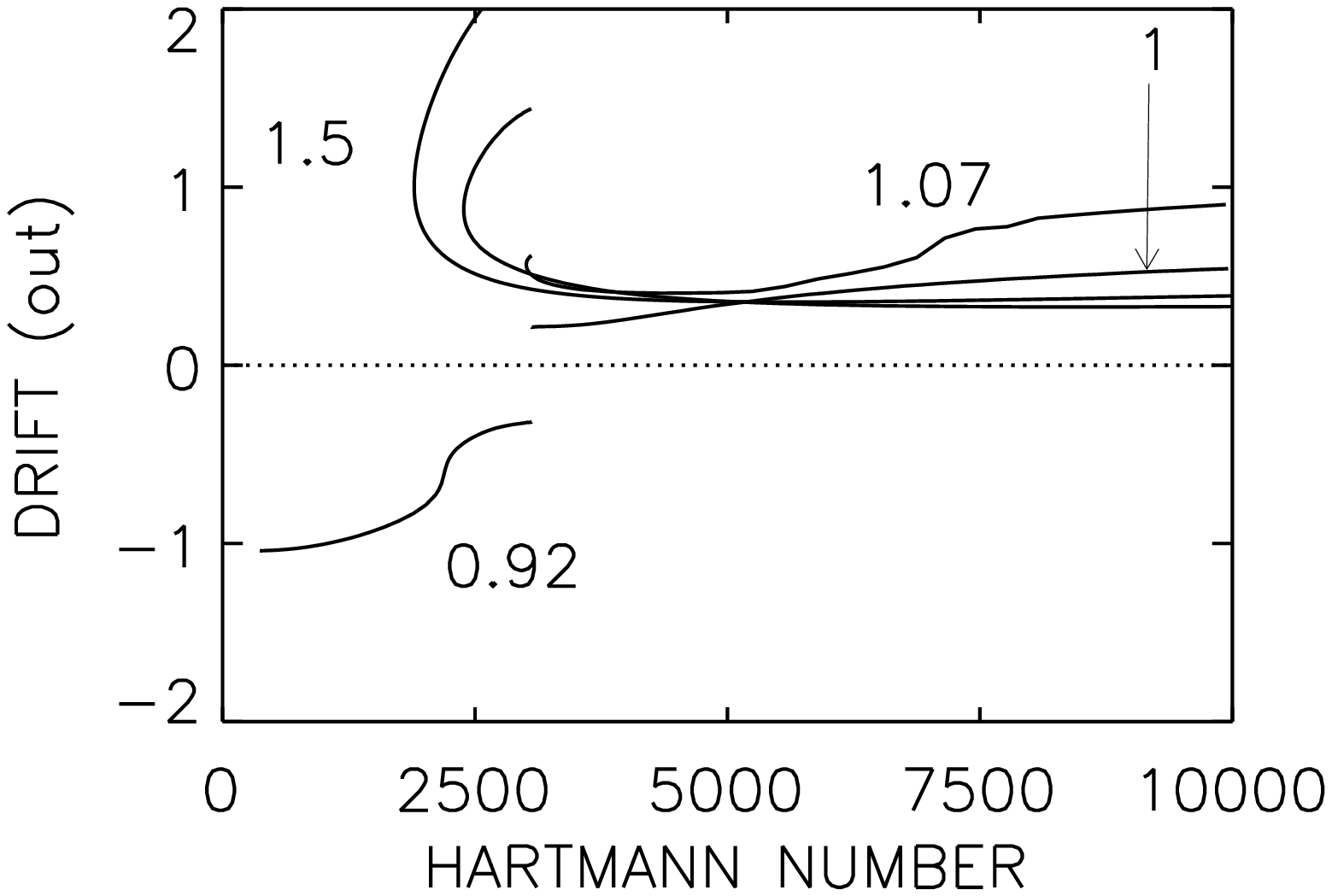}
     \caption{The same as in Fig. \ref{f8} but for the normalized drift frequency  $\omega_{\rm dr}/\Om_{\rm out}$ (see Eq. (\ref{drift})) in a narrow gap ($\rin=0.95$) for  $\rm Pm=10$ (left),  $\rm Pm=1$ (middle) and  $\rm Pm=0.1$ (right).  $\rin=0.95$, perfect-conducting boundaries.}
    \label{f88}
 \end{figure}

{Figure  \ref{f88} shows  the  behavior of the  azimuthal migration of the nonaxisymmetric vortices as more diverse.  It is  striking that for large $\Pm$ and/or for sub-rotation often $|\omega_{\rm dr}|\simeq \Om_{\rm out}$. 
}
From Fig. \ref{f88} we also find 
  that the nonaxisymmetric instability pattern for sub-rotation nearly corotates with the outer cylinder (as it is observed for AMRI\cite{R14,S14}) 
  for all magnetic Prandtl numbers. For  $\Pm \gsim 1$  the drift frequencies  for super-rotation are also negative  so that  their  magnetic pattern azimuthally  migrates in die direction of the rotation. 
 
For smaller magnetic Prandtl numbers, however,  for rigid rotation and for super-rotation the pattern counterrotates. The azimuthal migration  of the linear solutions directly reflects  the actual value of the  magnetic Prandtl number. This is also true for  a rigidly rotating pinch. After the results plotted in Fig. \ref{f88} its pattern corotates with the outer cylinder for small $\Pm$ and it  counterrotates with the outer cylinder for large  $\Pm$. Hence, numerical simulations  with magnetic Prandtl number unity for flows  with  vanishing or positive  shear  may easily lead to  results which are { not} representative for the solutions with smaller $\Pm$. On the other hand, rotation laws with negative shear do not show that sensitivity to the $\Pm$-value (see also Fig. \ref{f71}, below).

\section{Wide gap }

\subsection{Instability map}
\begin{figure}[htb]
     \includegraphics[width=5.4cm]{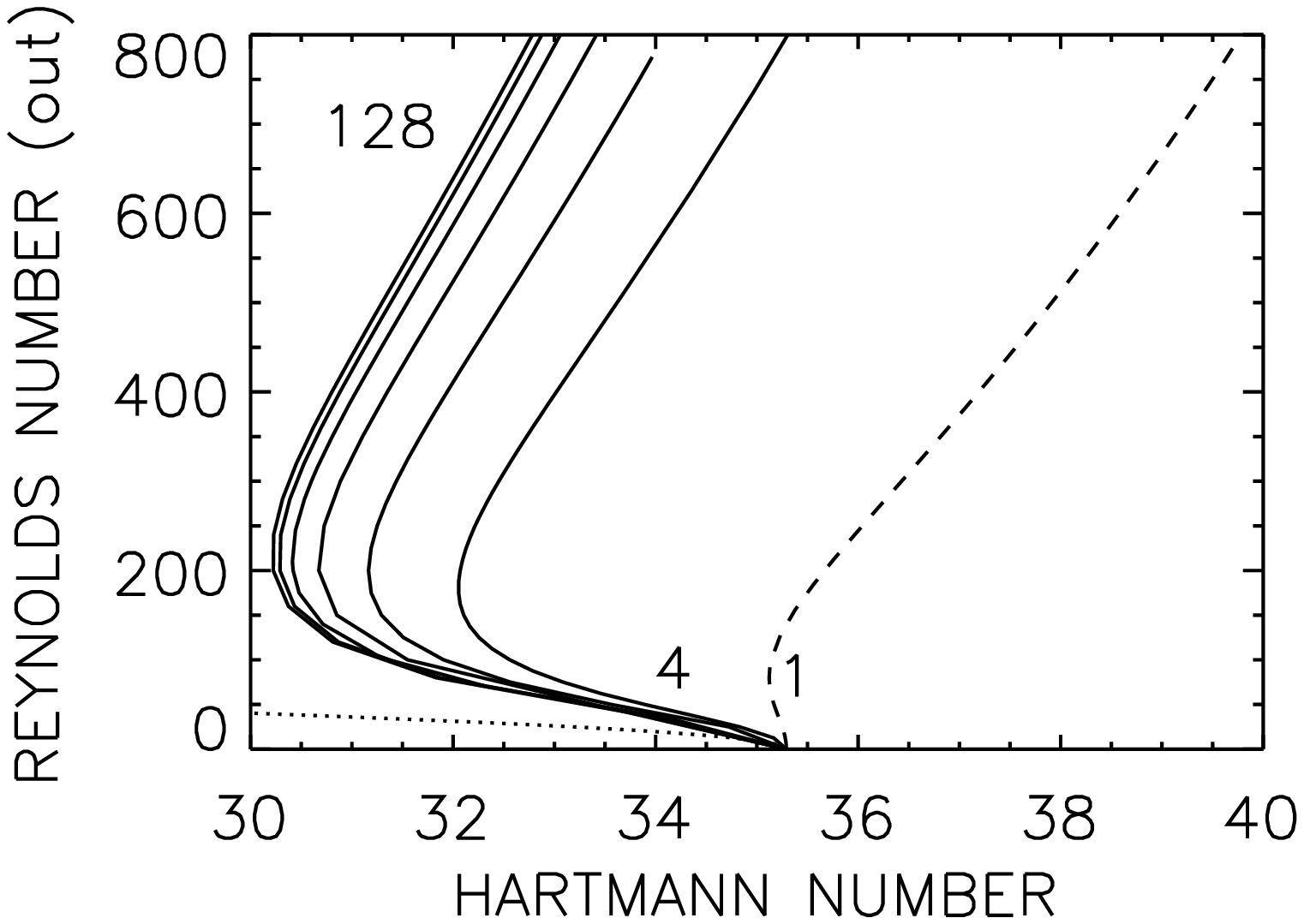}
     \includegraphics[width=5.4cm]{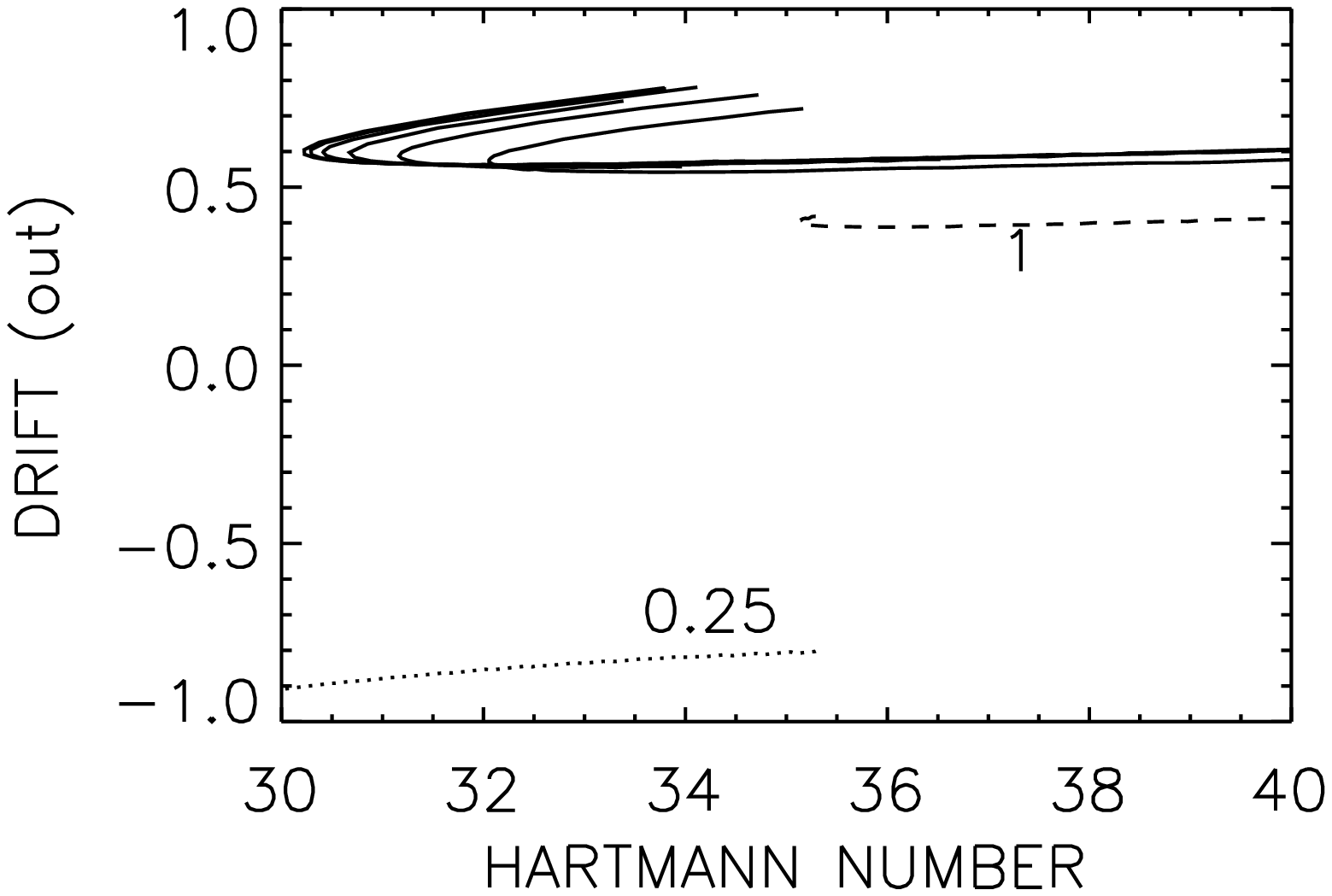}
     \includegraphics[width=5.4cm]{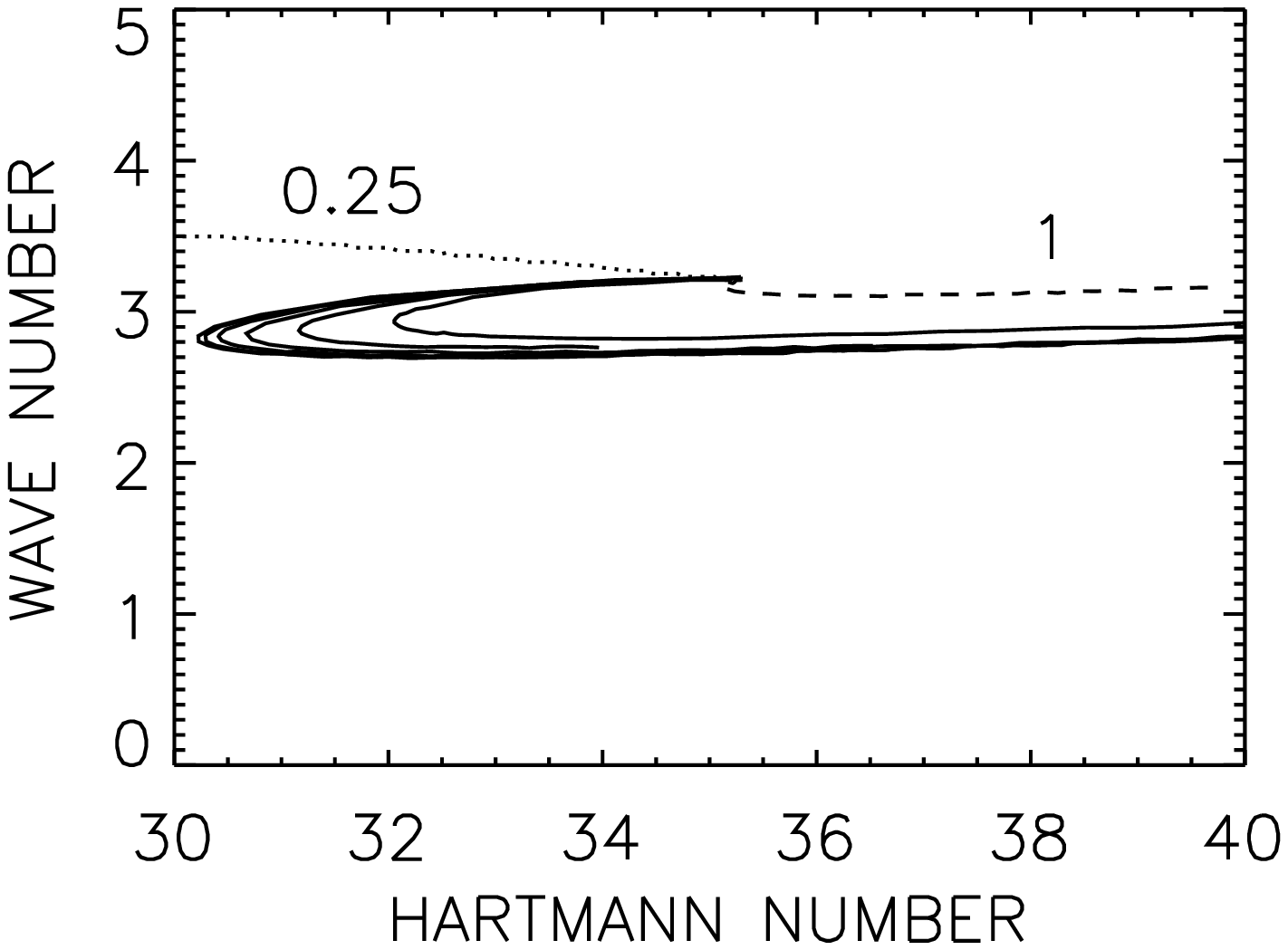}
 \caption{Left panel: instability  map  for   super-rotation in a wide gap, the lines are marked with   $\mu$. The dashed curve belongs to rigid rotation while  for reasons of comparison the dotted curve shows the result for $\mu=0.25$ (Rayleigh limit). Middle panel:  drift rates $\omega_{\rm dr}/\Om_{\rm out}$.  Right panel: the vertical wave numbers are nearly uniform. $\rin=0.5$, $\Pm=10^{-5}$, perfect-conducting cylinders.}
    \label{f71}
 \end{figure}

For a wide gap with $\rin=0.5$ and also in advance  to   a possible laboratory experiment Figs. \ref{f71}   give the eigenvalues for marginal instability, the wave numbers and the drift rates  for a fluid  with the magnetic Prandtl number of $\Pm=10^{-5}$  (liquid sodium). The characteristic Hartmann number for conducting boundaries and for resting cylinders is $\Ha_{\rm Tay}=35.3$  (see Fig. \ref{rigid}). From now on Reynolds numbers and drift rates are  related   to the rotation rate of the {\em outer} cylinder.
 
 Figure \ref{f71} (left panel) shows the curves of marginal instability for many rotational laws with $\mu$ between 1 and 128. The subcritical excitation of the current-driven instability by super-rotation is much stronger than for rigid rotation (dashed curve). If the Reynolds number is formed with the outer rotation rate, i.e. $\Rey_{\rm out}=\mu \Rey$,  then the  curves for strong shear are converging.  Hence, a minimum Hartmann number  of order 30 exists which cannot further be reduced by  steeper rotation laws. It is also clear that for $\Rey_  {\rm out}>200$ the differential rotation starts simply  to suppress the nonaxisymmetric instability. The middle panel of the Fig. \ref{f71} shows that the azimuthal drift of the instability pattern has the same (positive) sign for rigid rotation and super-rotation and the opposite (negative) sign for sub-rotation. The idea is thus supported that the phenomenon of subcritical excitation, i.e.  ${\rm d}  \Rey/{\rm d}\Ha<0$, for slow rotation with vanishing or positive shear is a common double-diffusive phenomenon which disappears for $\Pm=1$. On the other hand the known lowering of the critical Hartmann number and the negative   drift values by  sub-rotation   exists for {\em all} $\Pm$ (see Fig. \ref{f9}).  Note that for $\omega_{\rm dr}/\Om_{\rm out}= -1$ the magnetic pattern strictly corotates with the outer cylinder.
In contrast, the magnetic  pattern for {\em all} super-rotation laws including uniform rotation  migrates opposite to the sense of rotation corresponding to the behavior of the drift rates in the narrow gap. For small $\Pm$ by choice of the rotation rate  of the outer cylinder one can obtain all sorts of migration   of the magnetic pattern between corotation and counterrotation.

 The axial wave numbers $k$ are plotted in the right panel of Fig. \ref{f71}. They all show similar values. Let $\delta z$ be the characteristic vertical scale of a cell. Then from the definitions follows
 \beg
 \frac{\delta z}{D}\simeq \frac{\pi}{k}\sqrt{\frac{R_{\rm in}}{D}}
 \label{k}
 \ende
 so that $\delta z/D\simeq  \pi/k$  forms the relation between the cell size and the wave number for $\rin=0.5$.  Hence, the data  with $k\simeq \pi$ lead to  $\delta z\simeq D$ what means that the Tayler cells approximately form a circle in the $R$-$z$ plane. In this general formulation the  result  does  not depend on the gap width.

 \subsection{Growth rates}
 
\begin{figure}[htb]
    \includegraphics[width=8.0cm]{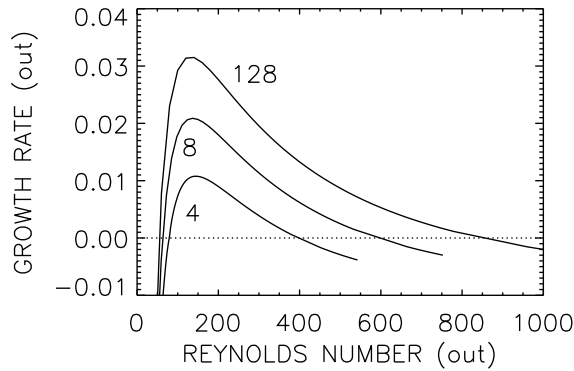}
 \caption{ Growth rates normalized with $\Om_{\rm out}$ for $\Ha=33$  with $\mu=4,8$,  and 128. $\rin=0.5$,  $\rm Pm=10^{-5}$, perfect-conducting cylinders.}
    \label{f74}
 \end{figure}

The subcritical excitation of the TI for rotation laws with positive shear only exists for sufficiently slow rotation ($\rm Mm\ll 1$) so that  the instability only  grows very slowly as   Fig. \ref{f74} demonstrates for super-rotation. The growth rates are normalized with the outer rotation rate. Then the maximum growth rates always occur for the same Reynolds number.  For the upper curve of the plot one finds that $\omega_{\rm gr}\simeq 0.03  \Om_{\rm out}$ for $\Rey\simeq 130$ so that the exponential growth time is $\tau_{\rm gr}\simeq 10 R_{\rm out}^2$  in seconds when $R_{\rm out}$ is measured in cm (it is $ \nu=7 \cdot 10^{-3}$ cm$^2$/s for liquid sodium). 
  If the expression 
 \beg 
 \omega_{\rm gr}= \Gamma \frac{B_{\rm out}^2}{\mu_0\rho\eta}
 \label{growth}
 \ende 
 of the growth rate is adopted (with $\Gamma$ as a dimensionless numerical factor which only depends on the gap width and the magnetic Prandtl number),   which has been derived for the nonrotating pinch and which has been experimentally realized \cite{RS10,S12}, then  for the strongest super-rotation the Fig. \ref{f74} gives $\Gamma\simeq 10^{-3}$ which very well fits  the theoretical  results for the resting container. This value  certainly  increases for wider gaps but the exponential growth time  of the instability for slow super-rotation will hardly be shorter than that for the resting pinch.

 \section{Summary and outlook}
\begin{figure}[htb]
 \includegraphics[width=8.0cm]{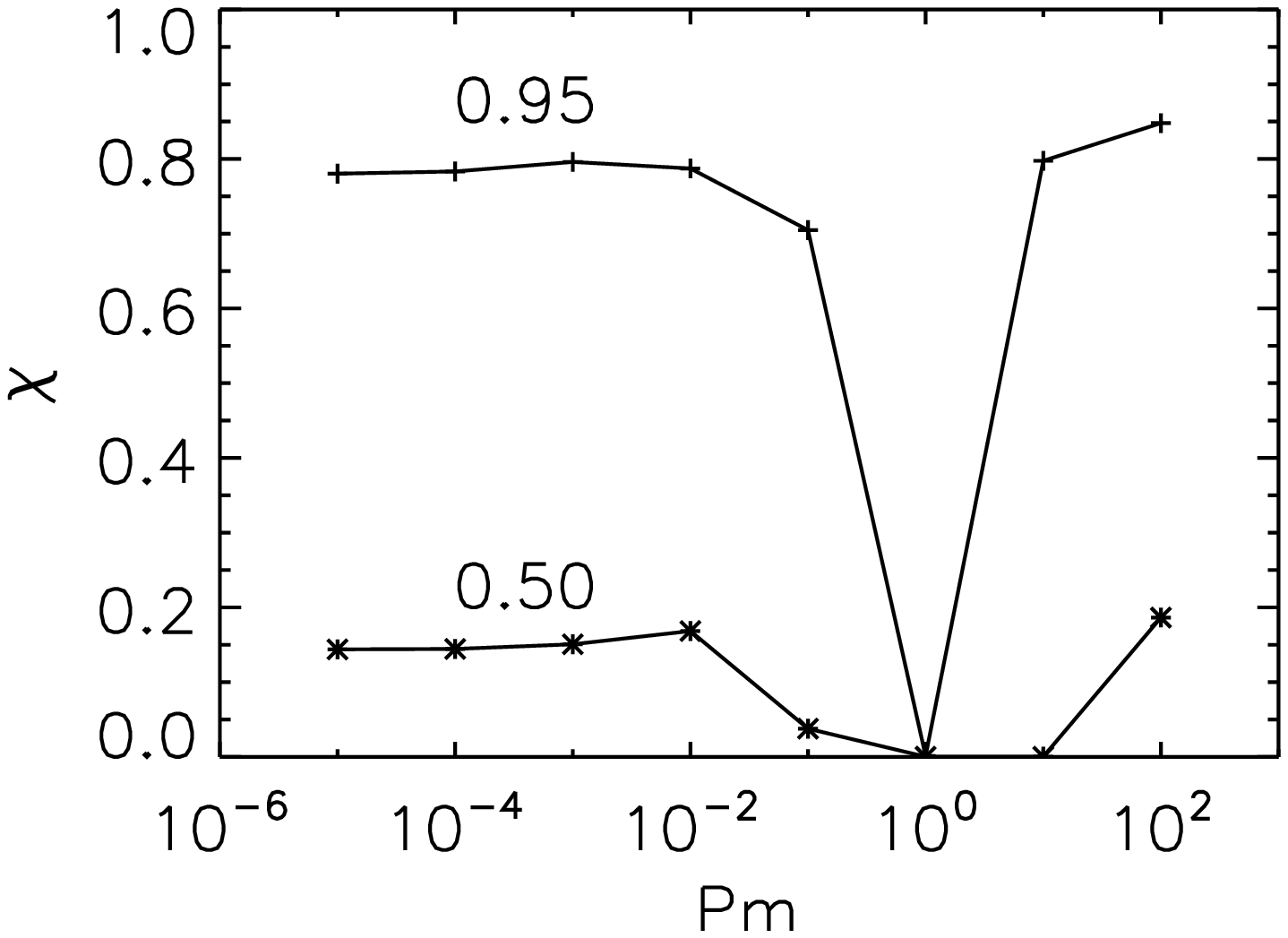}
 \includegraphics[width=8.0cm]{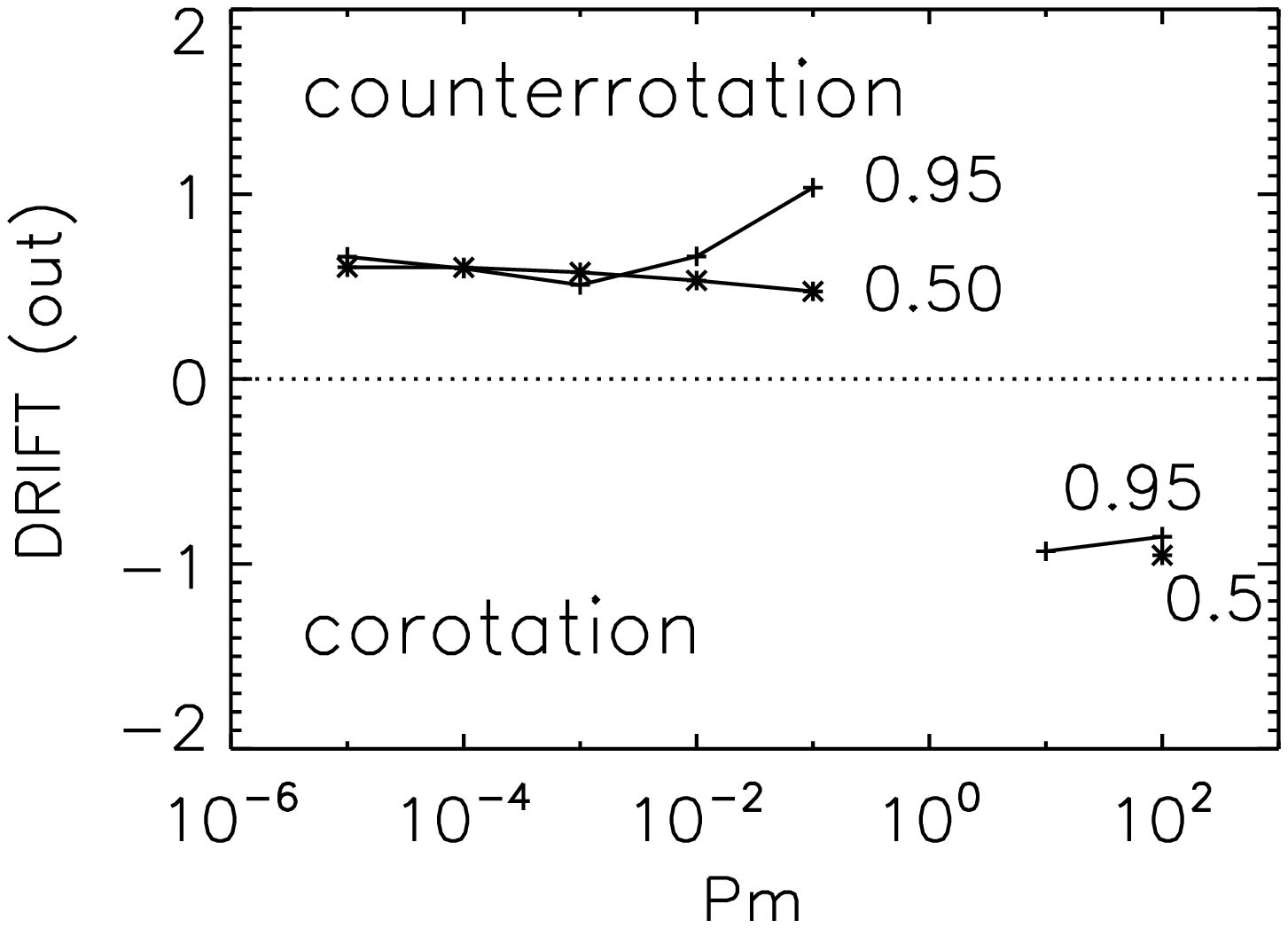}
     \caption{The  maximal reduction of the Hartmann number  (left) and the related  drift frequencies $\omega_{\rm dr}/\Om_{\rm out}$ (right)  as function of the magnetic Prandtl number. For $\Pm\gg 1$ and for $\Pm\ll 1 $ the subcritical excitation measured by (\ref{chi}) is very similar but the signs of the azimuthal migration of the instability pattern differ. The instability patterns drift in the rotational direction (corotation) or  opposite  (counterrotation).  Note the strong influence of the gap width on the Hartmann number reduction. The curves are marked with $\rin$, $\mu=\to \infty$, perfect-conducting boundaries.}
    \label{f72}
 \end{figure}

A rotating pinch with a homogeneous axial electric-current has been considered where the cylindric bounding walls rotate with $\mu\geq 1$, i.e. the outer cylinder rotates with the same rotation rate or rotates  faster than the inner cylinder. A linear perturbation theory fixes the critical Hartmann numbers for which the system becomes marginally unstable. The surprising result is the occurrence of a double-diffusive instability which only exists for $\nu\neq \eta$. For slow rotation with $\Om_{\rm out}\ll\Om_{\rm A}$ the excitation of the nonaxisymmetric perturbations becomes subcritical, i.e. the critical Hartmann number for rotation is smaller than without rotation. The effect is rather weak  for rigid rotation but it is remarkably strong  for super-rotation. It is numerically shown that for steeper and steeper  rotation laws the series of minimum Hartmann numbers converge to a total minimum   $\rm Ha_{Min}$  for  $\mu\to \infty$ (approaching resting inner cylinders). The resulting normalized lowering  of $\Ha$
\beg
\chi= \frac{\Ha_{\rm Tay}-\Ha_{\rm Min}}{\Ha_{\rm Tay}}
\label{chi}
\ende
 depends on the magnetic Prandtl number $\Pm$.  It vanishes  for $\Pm=1$ and takes similar values for very large and for very small $\Pm$. Relaxations of the the critical Hartmann number  of order 20\% ($\rin=0.5$) and 80\% ($\rin=0.95$) exist for $\Pm\neq1$ (Fig. \ref{f72}, left panel). Note also that the excess (\ref{chi}) grows for decreasing gap width.
 
It is also shown for a narrow gap and for small magnetic Prandtl number that  in the area of the instability map where the subcritical excitation exists for super-rotation the form of the lines of marginal instability only weakly depends on the radial profile of the azimuthal magnetic field. { For finite Reynolds number the lines of marginal instability in  Fig. \ref{f7} for the two different 
profiles (\ref{basic})  and  (\ref{basic2}) are  very close together}. In this domain of the  map the radial distribution of the axial electric-current seems to be unimportant. Obviously,   the  energy  provided by the differential rotation is large enough  to maintain the instability while the magnetic field is only needed as a catalyst.  This is a numerical finding which   implies that among all other radial profiles also the  profile with $B_R\propto 1/R $  (no current within  the fluid)  leads to a nonaxisymmetric instability for super-rotation. Again, however, this phenomenon  disappears for $\Pm=1$ revealing its double-diffusive character. 

The instability for $\Pm\gg 1$ and for $\Pm\ll 1 $ differs in another respect. For sub-rotation we always find that the pattern migrates for all $\Pm$ in positive direction of the azimuthal coordinate $\phi$. For solid-body rotation and for super-rotation there is, however,  a strong influence of the magnetic Prandtl number on the azimuthal migration of the perturbation patterns.  For  $\mu\geq 1$ the pattern counterrotates for small $\Pm$ while it corotates for $\Pm> 1$.   The $\Pm$-dependence of the  drift frequency $\omega_{\rm dr}$ disappears for  small and for large magnetic Prandtl numbers (Fig. \ref{f72}, right panel). The latter finding  is also true for a pinch with solid-body rotation.
 
A final  question is whether the described phenomena can be realized in the laboratory. The answer is yes and there is an interesting variety of possibilities.The following estimations are derived for liquid sodium  with $\Pm=10^{-5}$ as the fluid  conductor.  In Fig. \ref{f71}   for a container with $r_{\rm in}=0.5$ the characteristic Hartmann number for resting containers is 35.3 while the   smallest Hartmann number for (strong) super-rotation is 30.4.  The necessary current in the gap to produce a certain Hartmann number $\Ha$ is 
\beg
 I_{\rm fluid}= 5 \sqrt{\frac{(1-\rin)(1+\rin)^2}{\rin^3}}\sqrt{\mu_0\rho\nu\eta}\ \Ha
 \label{current}
 \ende 
(R\"udiger et al. 2013\cite{RKH13}) which does not depend on the physical size of the container.  For liquid sodium it is $\sqrt{\mu_0\rho\nu\eta}\simeq 8.2$ G$\cdot$cm  (for gallium  $\simeq 26$ G$\cdot$cm). To reach $\Ha\simeq 30$ one needs  an electric-current of  3.6~kAmp  flowing  through the sodium (11~kAmp for gallium). 

After Fig. \ref{f7} experiments are more interesting for  narrow gaps  with $\rin=0.95$.  A characteristic minimum Hartmann number is $O(1000)$ for super-rotation which needs  about 20~kAmp for its generation with liquid sodium. The Tayler instability with resting cylinders would, however,  require about 45 kAmp for its realization which seems to be much too high. Only the interplay with differential rotation allows this instability to observe.  A particular challenge for  such experiments  is that already for  a slight increase of this Hartmann number the electric-current through the sodium can be replaced by an axial current inside the inner cylinder. Both resulting  instabilities are very similar with respect to wave number and azimuthal drift. Such an experiment would easily demonstrate the surprisingly close relation  of AMRI and TI under the presence of differential rotation. 

As the instability phenomena presented in this paper all belong to the class of small magnetic Mach numbers the technical realization of the Reynolds numbers required by  the lines in  Figs. \ref{f7} and \ref{f71}   is also no problem.
Note, however, that the needed rotation rates of the cylinders do indeed depend on the physical size of the experiment.
\acknowledgments{This work was supported in
frame of the Helmholtz Alliance LIMTECH as well as by Deutsche Forschungsgemeinschaft
under SPP 1488 (PlanetMag). The anonymous referees are acknowledged for several suggestions to improve the paper.
}

\appendix
\section{A local approximation}

 The described effects have mainly been calculated with  boundary conditions  for  perfect-conducting cylinders. After Fig. \ref{f9} the negative slope  ${\rm d} \Rey/{\rm d} \Ha<0$ of the lines of marginal instability for { slow} rotation  
  also exists for models with vacuum conditions but  with reduced efficiency.  It is thus worthwhile to discuss in plane geometry  (for narrow gaps)   the result 
 \beg 
  {\Rey}^{*2}=\frac{1}{4}  \frac{((1+\Ha^{*2} m^{*2})^2 - 4 \Ha^{*4} m^{*2})(1+\Ha^{*2} m^{*2})^2}
          {\Ha^{*4} {\rm Ro}^2 m^{*2} - ({\rm Ro}+1) ( (1+\Ha^{*2} m^{*2})^2 - 4 \Ha^{*4} m^{*2}) }, 
 \label{disp}
 \ende 
 of  a local approximation  for $\Pm\to 0$    where $\Rey^*$, $\Ha^*$ and $m^*$ represent the slightly modified Reynolds number, Hartmann number and azimuthal wave number\cite{KSF14}. The Rossby number ${\rm Ro}=(1/2) {\rm d log} \Om/{\rm d log} R$ represents the differential rotation, it is positive for super-rotation and negative for sub-rotation. In contrast to all other quantities the Rossby number enters the   Eq. (\ref{disp}) with odd and even powers. In (\ref{disp}) the  { Rossby number  ${\rm Ro}$ must be independent of the radius $R$ which only allows to consider   Taylor-Couette flows} of $0.2<\mu< 2$ (for $\rin=0.5$). The latter (super-)rotation law  corresponds to $\Ro\simeq 1$ in a very good approximation.

We shall  show that all curves with $m^*>1$ in a $\Ha^*$-$\Rey^*$ plane for slow rotation show a subcritical behavior compared with the critical eigenvalue $\Ha^*_{\rm Tay}=1/\sqrt{m^*(2-m^*)}$ for $\Rey^*=0$, i.e. $\Ha^*(\Rey^*)<\Ha^*_{\rm Tay}$.
To this end  the function $Z=(1+\Ha^{*2} m^{*2})^2 - 4 \Ha^{*4} m^{*2}$  is defined so that (\ref{disp}) yields
\beg 
  Z=  \frac{4 \Rey^{*2} \Ha^{*4}\Ro^2 m^{*2}}
          { (1+\Ha^{*2} m^{*2})^2 +4 (\Ro+1) \Rey^{*2}}. 
 \label{disp1}
 \ende 
For rigid rotation ($\Ro=0$) only the  solution $Z=0$ exists which does not reflect the rotational  influence as given by   Fig. \ref{rigid}.   Obviously, the function $Z(\Rey^*)$ only vanishes  for $\Rey^*=0$ and it is positive-definite for finite $\Rey^*$ if -- as we  shall assume -- $\Ro>-1$. The above mentioned eigenvalue  $\Ha^*_{\rm Tay}$ forms the solution of $Z=0$. It only exists for $m^*<2$. The solution of  $Z=\delta$ with $\delta>0$ can thus be written as $\Ha^{*2}= \Ha^{*2}_{\rm Tay}+\varepsilon$ with unknown sign of 
$\epsilon$ which, without loss of generality,  can be assumed as small against $\Ha_{\rm Tay}^*$. Hence, from the definition of the function $Z$ follows
 \beg 
  \varepsilon= -\frac{\delta}{4 m^*}, 
 \label{disp2}
 \ende 
 so that always  $\varepsilon<0$. For $\Ro>-1$ it is thus  $\Ha^*(\Rey)<\Ha^*_{\rm Tay}$ for  negative and positive shear, i.e. the  excitation of the TI becomes always  subcritical  by the action of  any differential rotation. It is possible to demonstrate that this result does not change without the restriction to small $\varepsilon$. 
 
 The fact that  $Z=0$  { requires} $\Rey^*=0$ has the consequence that $\Ha^*=\Ha_{\rm Tay}^*$ does not appear as a solution of (\ref{disp}) for finite Reynolds numbers. Hence, the curves of marginal instability always remain in the subcritical domain with $\Ha^*(\Rey^*)<\Ha^*_{\rm Tay}$ and never reach    Hartmann numbers larger than   $\Ha^*_{\rm Tay}$. The typical  suppression  of the magnetic instabilities    by fast rotation (see Fig. \ref{f71}) is  thus not reflected by  the local relation (\ref{disp}) for inductionless fluids.


\end{document}